\documentclass[a4paper,12pt]{article}
\pdfoutput=1 

\usepackage{jheppub} 

\usepackage{tabularx,subcaption,fancyhdr,mathtools,physics}
\usepackage{bm}
\usepackage{csquotes}
\usepackage{float}
\usepackage{hyperref}
\usepackage{color,latexsym, array,multirow,url,verbatim, enumerate}
\usepackage[normalem]{ulem}
\DeclareUnicodeCharacter{2212}{-}
\hypersetup{colorlinks=true,linkcolor=blue,citecolor=magenta,filecolor=magenta,
	urlcolor=cyan}
\def \lspone{\widetilde\chi_1^0}

\def \lsptwo{\widetilde\chi_2^0}

\def\chonepm{\widetilde{\chi}_1^{\pm}}

\newcommand{\beq}{\begin{equation}}
\newcommand{\eeq}{\end{equation}}
\def\bea{\begin{eqnarray}}
\def\eea{\end{eqnarray}}

\newcommand{\bmt}{\bm{\theta}}
\newcommand{\bmx}{\bm x}

\newcommand{\calP}{\mathcal{P}}
\newcommand{\calD}{\mathcal{D}}
\newcommand{\calL}{\mathcal{L}}

\newcommand{\bbE}{\mathbb{E}}

\newcommand{\vect}[1]{\mathbf{#1}}
\newcommand{\mat}[1]{\mathbf{#1}}
\newcommand{\br}[1]{\mathopen{}\left(#1\right)\mathclose{}}

\newcommand{\g}{\,|\,}

\newcommand{\deriv}[2]{\frac{\partial{#1}}{\partial{#2}}}
\newcommand{\gaussian}[2]{\mathcal{N}\br{#1,#2}}
\newcommand{\gaussianx}[3]{\mathcal{N}\br{#1\g #2,#3}}

\title{\boldmath 
Exploring the BSM parameter space with Neural Network aided Simulation-Based Inference  
}

\author[a, b]{Atrideb Chatterjee,}
\author[c]{Arghya Choudhury,} 
\author[d]{Sourav Mitra,}
\author[c]{Arpita Mondal,}
\author[e]{and Subhadeep Mondal}

\affiliation[a]{Kapteyn Astronomical Institute, University of Groningen, Landleven 12 (Kapteynborg, 5419) 9747 AD Groningen, The Netherlands}
\affiliation[b]{Inter-University Centre for Astronomy and Astrophysics, Post Bag 4, Ganeshkhind, Pune 411007, India}
\affiliation[c]{Department of Physics, Indian Institute of Technology Patna, Bihar - 801106, India}
\affiliation[d]{Surendranath College, 24/2 M. G. Road, Kolkata, West Bengal - 700009, India}
\affiliation[e]{Department of Physics, SEAS, Bennett University, Greater Noida, Uttar Pradesh  -201310, India}
\emailAdd{atrideb.chatterjee1994@gmail.com}
\emailAdd{arghya@iitp.ac.in}
\emailAdd{hisourav@gmail.com}
\emailAdd{arpita\_1921ph15@iitp.ac.in}
\emailAdd{subhadeep.mondal@bennett.edu.in}

\abstract{Some of the issues that make sampling parameter spaces of various beyond the Standard Model (BSM) scenarios computationally expensive are the high dimensionality of the input parameter space, complex likelihoods, and stringent experimental constraints. In this work, we explore likelihood-free approaches, leveraging neural network-aided Simulation-Based Inference (SBI) to alleviate this issue. We focus on three amortized SBI methods: Neural Posterior Estimation (NPE), Neural Likelihood Estimation (NLE), and Neural Ratio Estimation (NRE) and perform a comparative analysis through the validation test known as the \textit{ Test of Accuracy with Random Points} (TARP), as well as through posterior sample efficiency and computational time. As an example, we focus on the scalar sector of the phenomenological minimal supersymmetric SM (pMSSM) and observe that the NPE method outperforms the others and generates correct posterior distributions of the parameters with a minimal number of samples. The efficacy of this framework is tested on 5 parameter pMSSM with Higgs and flavor physics data and its performance is compared with the MCMC method. We further add dark matter (DM) observables to make the task more challenging and consider a 9 parameter pMSSM. We observe that even though the efficiency factor drops, the amortized SBI method still produces faithful posterior distributions. SBI predicted points satisfying DM constraints are mostly bino-dominated upto $\sim$ 1.5 TeV, and are mostly wino-dominated within the 1.5 - 2 TeV range.}

\begin{document} 
\maketitle
\section{Introduction}
\label{sec:intro}
The biggest question currently plaguing the High Energy Physics (HEP) community is the nature of new physics beyond the Standard Model (BSM). There are various categories of new physics scenarios all of which are well motivated from different theoretical and phenomenological perspectives, such as, neutrino oscillation ~\cite{deSalas:2020pgw, NOvA:2019cyt}, possible existence of Dark Matter (DM) ~\cite{2009GReGr..41..207Z, 1937ApJ....86..217Z, Jungman:1995df}, hierarchy problem~\cite{SUSSKIND1984181, PhysRevD.14.1667}, CP violation~\cite{BaBar:2015gfp, PhysRevD.108.012008, Belle:2022uod,Miao:2023cvk},  to name a few. 
There exist different classes of new physics models that can account for these shortcomings of the SM. These models introduce handful of new parameters and unless they are well constrained, prediction of the experimental observables would not be very unique. So, a major objective of phenomenological studies is to
narrow down the possible values of these parameters as much as possible, using both  theoretical calculations and experimental observations. Scalar search is of major interest as far as experimental probe of BSM physics is concerned. A lot of BSM scenarios predict the existence of new scalars in addition to the 125 GeV Higgs \cite{CMS:2012qbp,ATLAS:2012yve}. 
A simple example of such extensions would be the two Higgs doublet models (2HDM)~\cite{Branco:2011iw}.
One can use the mass and coupling measurement of 125 GeV Higgs 
boson~\cite{ATLAS:2015yey} to constrain these scenarios effectively. 

Among the four 2HDM scenarios, the scalar sector of the type-II 2HDM model~\cite{Branco:2011iw} 
is similar to that of the minimal supersymmetric standard model (MSSM)~\cite{Martin:1997ns}. Amid different BSM possibilities, MSSM remains a favorite among theorists due to its rich phenomenological implications and elegant mathematical framework. However, unlike the 2HDM model, the low scale phenomenological MSSM (pMSSM)\cite{MSSMWorkingGroup:1998fiq} introduces 19 new parameters, which makes it a daunting task to probe this parameter space 
efficiently\footnote{Several phenomenological groups have analyzed the pMSSM 
parameter space in the context of LHC data, dark matter and precision 
experiments \cite{Barman:2024xlc, Bagnaschi:2017tru, GAMBIT:2017zdo, Barman:2016jov, Bhattacherjee:2015sga, Roszkowski:2014iqa}}. 
For long, the standard practice was to carry out phenomenological studies by choosing random representative points (or benchmark points) from the whole parameter space. However, as the parameter space shrinks, it gets more and more difficult to choose these points if one takes into account the constraints and experimental data from all particle physics experiments. Moreover, rather than just identifying `good' or `bad' points, it is more  effective to identify the region of parameter space that is most favored by the existing experimental data. Then, one can focus on that preferred region for any further impactful phenomenological studies, which makes the results more relevant.

In order to explore and sample the high-dimensional parameter spaces effectively,  one has to adapt sophisticated statistical techniques. These parameter inference methods can be broadly classified into two categories: (i) Likelihood-based, and (ii) Likelihood-free methods. As is obvious, the likelihood-based method first takes into account the model prediction of experimental observables in order to explicitly calculate the likelihood and then minimizes it to estimate the posterior distribution of the parameters. Traditional methods, such as Markov Chain Monte Carlo (MCMC), fall into this category. Although the methods are extremely successful, they have two main limitations: (i) the problem may not even have a tractable likelihood, and (ii) calculating the likelihood can be  computationally expensive. 
The latter requires state-of-the-art computational facilities. Deep learning methods have been used to address this issue \cite{Hammad:2024tzz,Binjonaid:2024jpm,Diaz:2024sxg}. Active learning has been used extensively to study new physics parameter space subjected to experimental data. It is an iterative method that can identify `good' and `bad' points lying on either side of a {\it decision boundary} \cite{Caron:2019xkx,Goodsell:2022beo}. Using Machine Learning (ML) methods, one can map a higher dimensional parameter space to a lower dimensional {\it latent space}. One can now probe the lower dimensional space to locate clusters of `good' points \cite{Mutter:2018sra,He:2022fxp,Baretz:2023mra}. ML-assisted algorithms with both regression and classification have been developed to sample parameter space \cite{Hammad:2022wpq, Baruah:2025nby}. Efficient algorithms like Hamiltonian Monte Carlo have also been used to sample likelihood \cite{Hollingsworth:2021sii}. Diffusion models can be used to accelerate MCMC effectively \cite{Hunt-Smith:2023ccp}. Multi-objective active search techniques have been utilized to good effect \cite{Diaz:2024yfu}. Evolutionary strategies can converge very quickly on the favored parameter space. An effort has been made to leverage this with remarkable efficiency \cite{PhysRevD.109.095040}. 

To mitigate these problems, a recent trend towards leveraging Machine Learning (ML) techniques for posterior inference has been proposed that completely bypasses the need to calculate the likelihood function explicitly. These likelihood-free inferences are often called Simulation-Based Inference (SBI) as it requires a simulator for calculating observables from parameters (discussed in detail in Section~\ref{sec:method}). A subclass of these methods is known as amortized for the reason that once trained, the methods can infer the posterior distribution of the model parameters across various observations without training it on new data for each new observation \cite{2018JCoPh.366..415Z, Hermans_2019, papamakarios2018fastepsilonfreeinferencesimulation, papamakarios2019neural}. 

In the context of HEP, the use of neural-network based SBI frameworks is recently gaining attention~\cite{Brehmer:2020cvb,Bahl:2024meb, Mastandrea:2024irf, ATLAS:2024jry, Morrison:2022vqe, Ghosh:2025fma, Amram:2025vqw, ATLAS:2025clx,CMS:2025zue,DeLuca:2025ruv,Sluijter:2025isc,Acosta:2025lsu,Desai:2025mpy,Shyamsundar:2025eht, Bhattacharjee:2025ofd,Morandini:2023pwj}. However, its full potential and varied applications are yet to be explored. For example, SBI framework has been used to constrain the particle interaction at the Large Hadron Collider (LHC) via simultaneous estimation of Wilson co-efficient from di-boson ($W^\pm Z$)~\cite{Bahl:2024meb} and di-higgs ($hh$)~\cite{Mastandrea:2024irf} production  and extracting the likelihood ratio over the relevant parameter space in the effective theory extension of the SM (SMEFT) framework. 
The ATLAS Collaboration has recently explored the off-shell Higgs boson coupling measurements in the four lepton final states using Neural SBI framework by considering issues like incorporation of a large number of nuisance parameters in the analysis, quantification of the uncertainty from a limited amount MC generated events and implementation of NN to produce robust likelihood ratios and confidence intervals from the LHC data \cite{ATLAS:2024jry}. 
The authors in Ref.\cite{Morrison:2022vqe} have utilized the Sequential Neural Ratio Estimation (SNRE) algorithm to sample the parameter space of pMSSM, focusing on the dark matter and SUSY interpretation of the anomalous muon (g-2) with light sleptons. However, the SNRE method is not amortized.
In this work, we subject the pMSSM parameter space to two scenarios. First, we choose the pMSSM model considering 5 free parameters only (pMSSM5) and sample the parameter space according to precisely measured Higgs coupling strengths, its mass, and some flavor observables. We use this scenario to showcase the effectiveness of amortized SBI methods in generating posterior distributions of the BSM parameter space. In addition to that we apply collider constraints on the SUSY particle masses. Furthermore, we consider the pMSSM scenario with 9 free parameters (pMSSM9) and evaluate the efficiency of the SBI approach in exploring the viable pMSSM9 parameter space that satisfies current dark matter constraints along with the Higgs and flavor physics observables as considered in pMSSM5 scenario.


Verifying the accuracy of posterior inference obtained through SBI methods is not straightforward and is still an open issue. Typically, the accuracy of the estimated posterior is evaluated using coverage probability tests, which are necessary but not sufficient. Very recently, \cite{2023PMLR..20219256L} have introduced ``Tests of Accuracy with Random Points'' (TARP)  method to check the  accuracy of the estimated posterior and demonstrated that their approach  is both necessary and sufficient to validate the accuracy of a posterior estimator. We implement this new method to check 
the credibility of the posterior sample obtained by the  NPE, NLE, and NRE methods. It may be noted that the TARP test has been used in the literature mostly in the context of cosmological/astrophysical analyses \cite{Hahn:2023udg,Zhong:2024qpf}.

This article is organized as follows: In Section~\ref{sec:method}, we discuss various SBI methods and emphasize the necessity of the TARP test for validation. Section~\ref{sec:result_pmssm} covers the observables, parameters, and collider constraints on sparticles, along with the sample preparation process for SBI methods. Then, we present and compare the results obtained with different SBI methods in this section. We also compare the results coming from two methods - SBI and MCMC in Section~\ref{sec:result_pmssm}. We show how efficiently SBI method works for this model with more parameters and DM constraints in Section~\ref{sec:dm}. Here, we discuss the impact of DM constraints on the pMSSM parameter space. Finally, we summarize in Section~\ref{sec:summary}.

\section{Methods}
\label{sec:method}

This section provides a brief overview of the mathematical framework behind amortized SBI and its various methodologies. Additionally, we describe the TARP test in detail, a  necessary and sufficient validation test to ensure that the resulting posterior distribution accurately reflects the underlying true distribution.

\subsection{Simulation Based Inference}

All the amortized SBI methods start with Bayes' theorem to find the model parameters $\bmt$ given a set of observations $\bmx$, i.e., $\calP \left( \bmt \vert \bmx \right)$
\begin{equation}\label{eqn:bayes}
    \calP \left( \bmt \vert \bmx \right) = \frac{\calP \left( \bmx \vert \bmt \right) \ \calP \left( \bmt \right)}{\calP(\bmx)} = \frac{\calP \left( \bmx \vert \bmt \right) \ \calP \left( \bmt \right)}{\int \calP \left( \bmx \vert \bmt' \right) \ \calP \left( \bmt' \right) d \bmt'}
\end{equation}
where $\calP \left( \bmx \vert \bmt \right)$ is commonly known as the likelihood, , $\calP \left( \bmt \right)$ is called the prior and $\calP \left( \bmx \right)$ is the evidence. Note that the main goal of any SBI method is to obtain the posterior probability $\calP \left( \bmt \vert \bmx \right)$.

As already mentioned, the primary distinction between the traditional likelihood-based parameter estimation approach and SBI techniques is that the 
SBI methods do not explicitly calculate the likelihood function to perform inference. Instead, they use a deep neural network to directly estimate posterior distribution $\calP(\bmt|\bmx)$ (Neural Posterior Estimation or NPE) \cite{papamakarios2018fastepsilonfreeinferencesimulation, greenberg2019automatic} or likelihood $\calP(\bmx|\bmt)$ (Neural Likelihood Estimation or NLE) \cite{papamakarios2019neural} or the likelihood ratio (Neural Ratio Estimation or NRE)~\cite{Hermans_2019}. 
In NRE method, we estimate the quantity $r(\bmx, \bmt)$ defined as $\frac{\calP \left( \bmx \vert \bmt \right)}{\calP (\bmx)}$ over the $\{ \bmt, \bmx \}$ pairs in the training data set $\calD_{\rm train}$. On the other hand, in NPE (NLE) method, we use a neural-density estimator $q_{\phi}(\bmt|\bmx)$  ($q_{\phi}(\bmx|\bmt)$) for the target distribution $\calP(\bmt|\bmx_{o})$. Here, $\phi$ denotes the set of weights and biases of the neural network, and $\bmx_{o}$ is the true observation. Therefore, the main goal of any SBI method is to vary $\phi$ to get the optimized parameter set $\phi^*$ to achieve
\begin{equation}
	q_{\phi^*}(\bmt|\bmx) \simeq \calP(\bmt|\bmx)
	\label{eqn:optimized_q}
\end{equation}

\begin{figure}[!htb]
	\centering
	\includegraphics[width=1.0\textwidth]{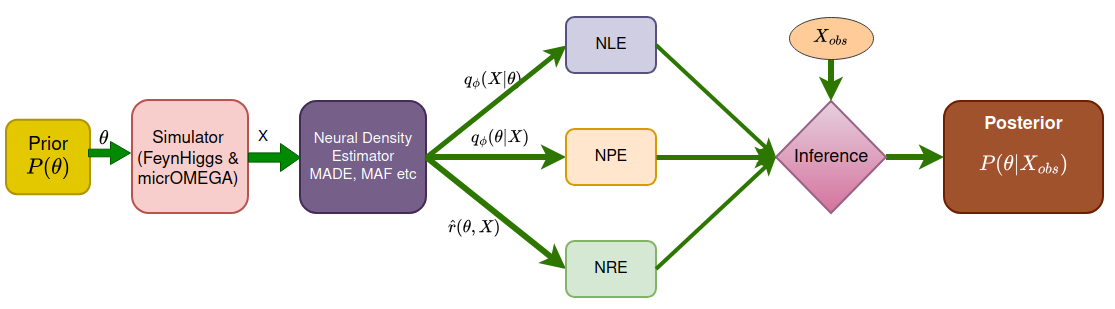}
	\caption{The flowchart of all the three SBI algorithms i.e., NPE, NLE and NRE is displayed in this figure.}
	\label{fig:sbi_method}
\end{figure}

In the following sections, we briefly review these three SBI methods along with their mathematical backgrounds, advantages, and limitations. We also describe one of the popular methods to calculate the estimator $q_{\phi}(\bmx|\bmt)$. The general flowchart of all three SBI methods  is illustrated in Figure.~\ref{fig:sbi_method}.

\subsubsection{Neural Posterior Estimation}
\label{sec:npe}
In this approach, one directly trains to learn the posterior distribution $\calP(\bmt \vert \bmx)$ by optimizing the loss function defined as negative of the {\it expectation} value of $\log \hat{\calP}(\bmt|\bmx)$ in the training dataset~\cite{papamakarios2018fastepsilonfreeinferencesimulation, greenberg2019automatic} i.e., 
\begin{equation}\label{eqn:Lnpe}
    \begin{aligned}
    \calL_{\rm NPE} &:= -\bbE_{\calD_{\rm train}}\log \hat{\calP}(\bmt|\bmx)\\
    &= -\bbE_{\calD_{\rm train}}\log \left[\frac{p(\bmt)}{\tilde{p}(\bmt)}q_{\phi}(\bmt|\bmx)\right],
    \end{aligned}
\end{equation}
where $\hat{\calP}(\bmt|\bmx)$  is the product of
a neural network output $q_{\phi}(\bmt|\bmx)$ and a weighting factor taken as the ratio of the assumed prior $p(\bmt)$ to the
proposal prior $\tilde{p}(\bmt)$.  The proposal prior is defined as the distribution of $\bmt$ present in the training dataset $\calD_{\rm train}$, while the assumed prior is an experimental design choice representing a priori knowledge of the global distribution of $\bmt$. 

\subsubsection{Neural Likelihood Estimation}
\label{sec:nle}
NLE~\cite{papamakarios2019neural} uses neural networks to estimate the likelihood function by optimizing the loss function given by
\begin{equation}\label{eqn:Lnle}
    \calL_{\rm NLE} := -\bbE_{\calD_{\rm train}}\log q_{\phi}(\bmx|\bmt).
\end{equation}
Once optimized over $\phi$, the trained $q_{\phi^*}(\bmx|\bmt)$
can be multiplied with $p(\bmt)$ to obtain $\hat{\calP}(\bmt|\bmx)$ i.e. $\hat{\calP}(\bmt|\bmx) \propto p(\bmt) q_{\phi^*}(\bmx|\bmt)$. 
One advantage of the NLE method over NPE (as is obvious from comparing equation \ref{eqn:Lnle} with \ref{eqn:Lnpe}) is that it does not require the knowledge of the analytical expression for $\tilde{p}(\bmt)$.
It is worth pointing out that in this method one calculates the likelihood for every parameter set only while preparing the dataset but not at the time of training or testing.

\subsubsection{Neural Ratio Estimator}
\label{sec:nre}
In this method, one first defines the likelihood ratio~\cite{Hermans_2019}
\begin{equation}\label{eqn:NRE}
    r(\bmx, \bmt) = \frac{\calP \left( \bmx \vert \bmt \right)}{\calP (\bmx)} = \frac{\calP (\bmx, \bmt)}{\calP(\bmx) \calP(\bmt)} = \frac{\calP^{\rm joint}}{\calP^{\rm marginal}}
\end{equation}
 In \citep{cranmer2015approximating}, the authors show that this equation ``can be used in a supervised learning setting to train a binary classifier $d_{\phi}(\bmx, \bmt)$ to distinguish joint samples $\{\bmx, \bmt\} \sim \calP^{\rm joint}$ with class label $y = 1$ from marginal samples $\{\bmx, \bmt\} \sim \calP^{\rm marginal}$ with class label $y = 0$". Once trained, the classifier can be written as
\begin{equation}\label{eqn:classifier}
    d_{\phi*}(\bmx, \bmt) =\frac{ \calP^{\rm joint}}{ \calP^{\rm joint}+\calP^{\rm marginal}}
\end{equation}
where the optimal parameters $\phi^*$ are obtained using an Adam optimizer~\cite{kingma2014adam} with binary cross-entropy loss of the form
\begin{equation}\label{eqn:BCE}
\calL_{\rm NRE}  = -\int \left[ \calP^{\rm joint} \ln{d_{\phi}(\bmx, \bmt)}+\calP^{\rm marginal}\ln\{1-d_{\phi}(\bmx, \bmt)\}\right]d\bmt d\bmx
\end{equation}
As is obvious from the equations \ref{eqn:NRE} and \ref{eqn:classifier},
the approximated ratio can then be obtained from
\begin{equation}
    r(\bmx,\bmt) = \frac{d_{\phi^*}(\bmx, \bmt)}{1-d_{\phi^*}(\bmx, \bmt)}.
\end{equation}
For estimating $r(\bmx,\bmt)$, we use the well established classifier \texttt{ResNET} \cite{DBLP:journals/corr/HeZRS15}. This is a widely used (across different domains of ML applications), publicly available neural network architecture. The main advantage of this model compared to other deep learning models is  the introduction of skip (residual) connections, which allow the input of a layer to bypass one or more subsequent layers and be added directly to their output. The skip connection is very efficient in alleviating the vanishing gradient problem which is a serious obstacle for very deep networks (often exceeding ~100 layers and comprising millions of parameters).

\subsection{Sequential vs. Amortized Inference in SBI}
\label{sec:snre_vs_sbi}
NPE, NLE, and NRE are amortized methods. Once trained, they can provide posterior or likelihood estimates for any new observation without additional simulations, making them particularly efficient for repeated inference tasks. Their main disadvantage is that they typically require a larger upfront simulation budget and training time to achieve accurate results, and they may spend computational effort on regions of the parameter space that are less relevant to the current observation.
On the other hand, Sequential Neural Ratio Estimation (SNRE) runs sequentially, focusing iteratively on the high-probability regions of the parameter space. This sequential strategy makes it highly simulation-efficient for estimating likelihood ratios or performing model comparison, especially in higher-dimensional spaces. It effectively concentrates computational effort on the most relevant regions, reducing wasted evaluations. However, SNRE is not amortized: it must be retrained for each new observation, which can be time-consuming if multiple datasets or repeated inference tasks are required. 

\subsection{Choosing the right method and validation test}
\label{sec:density_function}
The general guidelines for choosing one method over another are the following:
\begin{itemize}
	\item If the data is high-dimensional (e.g., in case of images) NPE may be a more viable choice than NLE or NRE.
	\item In some problems, the likelihood may be more straight-forward for neural learning than the posterior distribution. For such case, NLE is preferred over NPE or NRE.
	\item In case of NRE, one typically uses very well known neural-network architecture such as \texttt{ResNET} as mentioned above. It is a big advantage over NPE or NLE considering one has to spend a significant amount of time in choosing the suitable network for NPE and NLE. 
\end{itemize}

However, in reality, for most of the cases, one has to try all three SBI methods and only then depending on the performance one chooses one method over another.
Now we will discuss in more detail about the choice of estimator $q_{\phi}(\bmx|\bmt)$. As emphasized earlier, the main aim of the neural network will be to obtain $q_{\phi}(\bmx|\bmt)$ by varying $\phi$ such that $	q_{\phi^*}(\bmx|\bmt) \simeq p(\bmx|\bmt)$. With great advancement in neural network architecture in the last few years, the de-facto choices for the estimator become neural density networks such as Masked Autoregressive Flow (\texttt{MAF}) \citep{papamakarios2017masked}, Mixture density networks (\texttt{MDN}s)\citep{bishop1994mixture}, and Neural Spline Flow (\texttt{NSF})~\cite{durkan2019neuralsplineflows}.

In this work, we use \texttt{MAF}, which has demonstrated excellent performance across various general-purpose density estimation tasks. \texttt{MAF} starts from a base distribution, typically a Gaussian $\vect{z_0} \sim  \gaussian{\vect{0}}{\mat{I}}$ and applies a series of autoregressive functions $(f_{1}, f_{2}, ... f_{k})$ to obtain the target distribution $\vect{z_K}$,~\cite{Papamakarios_2018} i.e.,
\begin{equation}
	\bmx = \vect{z_K}
	\quad\text{where}\quad
	\begin{array}{l}
		\vect{z}_0 \sim \gaussian{\vect{0}}{\mat{I}}\\
		\vect{z}_k = f_k\br{\vect{z}_{k-1}, \vect{\bmt}}.\\
	\end{array}
	\label{eq:maf_gen}
\end{equation}
Here, every $f_{k}$ is a bijection mapping implemented by Masked autoencoder
for Distribution Estimation (MADE)~\cite{germain2015made} and is characterized by hyperparameters ``Number of Hidden Layer'' and ``Number of Transformer''.
\begin{equation}
	q^{\rm MAF}_{\phi}(\bmx|\bmt) =
	\gaussianx{\vect{z}_0}{\vect{0}}{\mat{I}}
	\prod_k{\abs{\det\br{\deriv{f_k}{\vect{z}_{k-1}}}}^{-1}}    
\end{equation}
The learnable parameters $\phi$ can then be obtained by optimizing the loss function
\begin{equation}\label{eqn:Lmdn}
	\calL_{\rm MAF} := -\bbE_{\calD_{\rm train}}\left[\log q^{\rm MAF}_{\phi}(\bmx|\bmt)\right],
\end{equation}
where the expectation $\bbE_{\calD_{\rm train}}$ is taken over all data-parameter pairs in the training set $\calD_{\rm train}$. Once optimized, $q_{\phi}^{\rm MAF}$ approaches the true likelihood function $\calP \left( \bmx \vert \bmt \right)$.

\subsubsection{Testing the validity of the SBI methods - TARP}
\label{sec:tarp}
The aim of any validation method for SBI is to quantitatively determine the accuracy of the estimated posterior distribution $\hat{\calP}(\bmt|\bmx)$ compared to the true posterior distribution $\calP(\bmt|\bmx)$. Very recently, the authors in Ref.~\cite{2023PMLR..20219256L} have proposed the TARP method to validate different SBI methods. The authors argued that this approach is necessary and sufficient to show that a posterior estimator is accurate. They also showed that this can detect inaccurate inferences in cases where other existing methods fail. Described below is the step-by-step procedure for the TARP test.

\begin{enumerate}
    \item One first separates the test dataset pair $\calD_{\rm test} = \left\{\bmt^{i}_{\rm test}, \bmx^{i}_{\rm test}\right\}^{i=N_{\rm test}}_{i=1}$ from the train data set $\calD_{\rm train}$. We make sure that the test dataset has never been used while training.
    \item Take a test point $\bmx^{i}_{\rm test}$, choose a random reference point $(\bmt^{i}_{r})$ from a Uniform distribution in the range $[0,1]$.
    \item Draw $N_{\rm p}$ points from the estimated posterior distribution $\hat{\calP}(\bmt|\bmx)$  i.e., \newline $\{ \bmt^{i, p}_{\rm predicted}\}^{p=N_{\rm p}}_{p=1} \sim \hat{\calP}(\bmt|\bmx^{i}_{\rm test})$.
    \item Next we calculate the euclidean distance between the reference point $\bmt^{i}_{r}$ and $\bmt^{i, p}_{\rm predicted}$ and denote this set by $\left\{d(\bmt^{i, p}_{\rm predicted}, \bmt^{i}_{r})\right\}^{p=N_{p}}_{p=1}$.  
    \item We also calculate distance $d(\bmt^{i}_{\rm test}, \bmt^{i}_{r})$ between true parameter ($\bmt^{i}_{\rm test}$) and the reference point. 
    \item Compute the fraction of the points $\left\{d(\bmt^{i, p}_{\rm predicted}, \bmt^{i}_{r})\right\}^{p=N_{\rm p}}_{p=1}$ with distance smaller than $d(\bmt^{i}_{\rm test}.  \bmt^{i}_{r})$. This fraction is called the credibility level defined as $1-\alpha$, where $\alpha$ is the confidence level. The higher the fraction, better is the estimate.
    \item Steps 2-6 are repeated for all the $N_{\rm test}$ points in the test dataset.
    \item Once the credibility level is determined for all the test data pairs, we then compute the histogram describing the coverage distribution of $\alpha$. The cumulative distribution of $\alpha$ multiplied with the bin width of the histogram is called the Expected Coverage Probability (ECP). For an accurate posterior estimator, $\rm ECP = 1-\alpha$, i,e, a straight line with $45^{0}$ inclination with the x-axis where x and y-axis represent the credibility level and expected coverage, respectively.
\end{enumerate}

In summary, the credibility level refers to the probability threshold used to define the credible interval or confidence level, indicating the degree of belief that the true parameter lies within this interval. On the other hand, the expected coverage is the number of times the true parameter value is actually contained within the credible intervals when averaged over many independent samples. We show the TARP test result in Section.~\ref{sec:result_tarp}.

\section{LtU-ILI framework for pMSSM5 with Higgs and flavor constraints}
\label{sec:result_pmssm}
In this section, we consider a pMSSM scenario with five free parameters (pMSSM5 model) to explore 
the Higgs sector. 
We outline the possible range of the parameters, constraints on sparticle masses imposed by LHC searches to date and identify the currently allowed parameter space for further exploration. 
The focus is primarily on the scalar sector, along with other key observables, such as flavor decay branching ratios. The scalar sector of pMSSM consists of two ${\rm SU(2)_L}$ doublets that result in two neutral CP even, one neutral CP-odd, and one charged Higgs states. Among the two CP-even states, the lighter one can be interpreted as the SM-like Higgs. Given the precise measurements of the corresponding mass and coupling strengths, it cannot have large mixing with the other CP-even state. One can use this to constrain the scalar sector parameters such as $M_A$, $\mu$, $\tan\beta$, and $A_t$. We take this approach coupled with direct search constraints of SUSY particles and flavor constraints for our analysis. Finally, we outline the preparation of the sample containing parameters with corresponding observables and present the results of our analysis.

We implement the SBI method using publicly available \texttt{LtU-ILI} (Learning the Universe Implicit Likelihood Inference) code~\cite{2024OJAp....7E..54H}. This package trains neural network based SBI models for estimating posterior distributions of the model parameters.
Additionally, the LtU-ILI code offers a variety of tools that users can utilize for testing and validating the trained models. We refer the interested readers to follow the Ref.~\cite{2024OJAp....7E..54H} for more details.
 

\subsection{Constraints on SUSY particles from collider searches}
\label{sec:constraint}
Both the ATLAS and CMS Collaborations  have conducted extensive searches for SUSY over the years. These searches cover a wide range of final states, looking for evidence of sparticles. Some key final states explored include jets + missing transverse energy (MET), leptons + jets + MET, multilepton final states, photons + MET, $b$-tagged jets + MET, and long-lived particles, etc. ATLAS and CMS have excluded gluino masses up to $\sim$ 2.0-2.3 TeV from various final states 
for lightest supersymmetric particle (LSP) $\lspone$ masses up to 600 GeV~\cite{CMS:2019zmd, ATLAS:2020syg, ATLAS:2021twp, CMS:2021beq, CMS:2019ybf, CMS:2020cpy, CMS:2022idi, CMS:2020bfa}. 
Similarly light squarks are also limited upto 1.25-1.85 TeV for a massless LSP~\cite{CMS:2019zmd,ATLAS:2020syg,ATLAS:2021twp,CMS:2021beq,ATLAS:2023afl,ATLAS:2022zwa}. 
Additionally, sleptons are excluded up to 700 GeV~\cite{ATLAS:2019lff, CMS:2020bfa}, wino-like chargino and second lightest neutralino upto $\sim$ 900-1450 GeV~\cite{ATLAS:2022zwa,ATLAS:2021yqv,CMS:2021cox,CMS:2022sfi}, 
higgsino-like charginos and heavy neutralinos upto 800 GeV~\cite{CMS:2022sfi,CMS:2024gyw}, assuming a massless bino type neutralino.
However, it is worth mentioning that if the LSP mass is above 500 GeV, there is no limit on wino-like chargino or sleptons. Pure higgsino-like chargino-neutralinos are also excluded upto 205-210 GeV 
\cite{CMS:2021edw, ATLAS:2021moa}. 
The long-lived charginos are also searched at the LHC and are excluded up to 610-884 GeV GeV~\cite{ATLAS:2022rme,CMS:2020atg} depending upon the lifetime. While performing the analysis, we maintain all these collider limits on sparticles.

\subsection{Observables}
\label{sec:observable}
As we have mentioned above that we mainly focus on the Higgs sector for our analysis. The discovery of the Higgs boson in 2012 by the ATLAS and CMS experiments at the LHC was a monumental achievement in particle physics, confirming the last missing piece of the Standard Model (SM). However, this discovery also opens the door to exploring Beyond the Standard Model (BSM) physics. We know that pMSSM can easily stabilize the Higgs mass against quantum corrections~\cite{MSSMWorkingGroup:1998fiq, Allanach:2004rh}. The tree-level Higgs mass depends on $\tan\beta$, the ratio of two vacuum expectation values (vev), and $M_A$, the mass of pseudoscalar Higgs. But there is a radiative correction to the tree level mass and this correction depends on many other pMSSM parameters like $A_t$, $\mu$, and soft SUSY breaking parameters. The Higgs decays also have QCD and electroweak corrections which depend on various SUSY parameters. 
Here we consider different observables like Higgs mass $m_h$, Higgs coupling strength for both the production mode ($\mu_i$) and decay channel ($\mu^f$), and the branching ratios of $B$-hadron decays. From the combined results provided by ATLAS and CMS analyses, we know that the SM-like Higgs boson mass is 125.09$\pm$0.21(stat.)$\pm$0.11(syst.) GeV~\cite{CMS:2012qbp,ATLAS:2012yve,ATLAS:2015yey}. But we know that there are theoretical uncertainties in the SUSY framework~~\cite{Allanach:2004rh} and due to that we consider Higgs mass as $125\pm1$ GeV. The signal strength for a production mode $i$ and a particular decay channel $f$ is defined as $\mu_i = \frac{\sigma_i}{(\sigma_i)_{SM}}$ and $\mu^f = \frac{\mathcal{B}^f}{(\mathcal{B}^f)_{SM}}$ respectively. Both the experiments, ATLAS and CMS have provided these $\mu_i$ and $\mu_f$ values. We consider the values provided by CMS collaboration~\cite{CMS:2020gsy}. We also consider three different flavor physics observables such as $\mathcal{B}r(B \rightarrow X_s \gamma)$~\cite{HFLAV:2019otj}, $\mathcal{B}r(B_s \rightarrow \mu^+ \mu^-)$~\cite{LHCb:2021vsc} and $\frac{\mathcal{B}r(B \rightarrow \tau \nu)}{\mathcal{B}r(B \rightarrow \tau \nu)_{SM}}$~\cite{Belle:2010xzn, Belle:2014eud, Belle:2015odw}. The best-fit and $1\sigma$ values of all these observables are mentioned in the Table.~\ref{tab:observable}.

\begin{table}[!htb]
\begin{center}
\begin{tabular}{||c|c||c|c||}
\hline\hline
\textbf{Observable} & \textbf{Best-fit $\pm~1\sigma$}  & \textbf{Observable} & \textbf{Best-fit $\pm~1\sigma$}\\ 
\hline\hline
 $m_h$ (GeV) & $125\pm 1.0$ & $\mu_{WH}$ & 1.46$^{+0.37}_{-0.35}$  \\
\hline
 $\mathcal{B}r(B \rightarrow X_s  \gamma)$ & (3.32 $\pm$ 0.15)$\times 10^{-4}$  & $\mu_{ZH}$ & 0.98$^{+0.31}_{-0.30}$ \\
 \hline
 $\mathcal{B}r(B_s \rightarrow \mu^+ \mu^-)$ & (3.09$^{+0.46 + 0.15}_{-0.43 - 0.11}$)$ \times 10^{-9}$ & $\mu^{ZZ}$ & 0.93$^{+0.10}_{-0.09}$ \\
 \hline
 $\frac{\mathcal{B}r(B \rightarrow \tau \nu)}{\mathcal{B}r(B \rightarrow \tau \nu)_{SM}}$ & 1.28 $\pm$ 0.25 & $\mu^{WW}$ & 1.20$^{+0.16}_{-0.15}$ \\
 \hline
$\mu_{ggH}$ & 1.04 $\pm$ 0.09 & $\mu^{bb}$ & 1.11$^{+0.20}_{-0.19}$ \\ 
\hline
$\mu_{VBF}$ & 0.75$^{+0.19}_{-0.17}$  & $\mu^{\tau\tau}$ & 0.80$^{+0.17}_{-0.16}$ \\
\hline
$\mu_{ttH}$ & 1.14$^{+0.21}_{-0.20}$  & $\mu^{\gamma\gamma}$ & 1.07$^{+0.12}_{-0.10}$ \\
\hline\hline
\end{tabular}
\caption{The best-fit along with $1\sigma$ value of fourteen observables
 considered in our analysis are mentioned here.}
\label{tab:observable}
\end{center}
\end{table} 

\subsection{Parameter space of pMSSM5 }
\label{sec:parameter}
There are 19 free parameters in  pMSSM, and not all of them have a direct or significant effect on the Higgs sector and flavor observables. For our analysis, we vary five of these parameters, which are most relevant while keeping others fixed at values decoupled from the rest. While we fix the parameters, we keep in mind the limits on sparticle masses coming from collider searches which are already discussed in Section.~\ref{sec:constraint}. To avoid the limits on squarks and sleptons, we decouple all the gluino, squark, and slepton masses by keeping them at 4 TeV. As we have mentioned for the LSP masses above 500 GeV, there is no limit on chargino-neutralino masses, we fix the LSP ($\lspone$) mass at 500 GeV. After that, we are left with wino mass parameter ($M_2$), Higgsino mass parameter ($\mu$), ratio of two vev ($\tan\beta$), pseudoscalar Higgs mass ($M_A$) and trilinear top coupling ($A_t$). The ranges of 
five parameters used for the scanning are mentioned in the Table.~\ref{tab:parameter}. We have also done a separate analysis for the nine dimensional pMSSM (pMSSM9) parameter space by additionally varying bino, gluino, squarks, sleptons mass parameters to satisfy the DM,  which is discussed in the Section~\ref{sec:dm}.

\begin{table}[!htb]
\begin{center}
\begin{tabular}{||c|c||}
\hline\hline
\textbf{Parameter} & \textbf{Range} \\ 
\hline\hline
 $M_2$ & 550-5000 \\
 \hline
 $\mu$ & 550-5000 \\
 \hline
 $\tan\beta$ & 1-60 \\
 \hline
 $M_A$ & 100-5000 \\
 \hline
 $|A_t|$ & 0-8000 \\
\hline\hline
\end{tabular}
\caption{Range of five free parameters considered for our analysis with pMSSM5 scenario. All the parameters mentioned here have unit GeV except $\tan\beta$. }
\label{tab:parameter}
\end{center}
\end{table} 

\subsection{Sample preparation for SBI}
\label{sec:data_file}
To sample the parameter space, we use the \texttt{FeynHiggs}~\cite{Heinemeyer:1998yj,Heinemeyer:1998np,Bahl:2016brp,Bahl:2018qog} package, and for calculating the branching ratios of $B$-hadron decays, we employ the \texttt{micrOMEGAs}~\cite{Belanger:2004yn,Belanger:2008sj,Belanger:2010gh,Belanger:2013oya,Belanger:2020gnr}. We initially perform a random scan of the entire parameter space to generate a sample file. This sample can be used for training purposes. However, to improve the efficiency of the prepared sample, we retain only those points that are consistent with all observables within the 3$\sigma$ limit of their best-fit values, as listed in Table~\ref{tab:observable}. We generated approximately $6.0\times 10^5$ samples randomly, of which only around $2\times 10^5$ samples remain after applying the constraints. As mentioned above, we calculate the observables value corresponding to each parameter set using  \texttt{FeynHiggs} and \texttt{micrOMEGAs}. To generate $6.0\times 10^5$ number of samples through those packages running parallel in 4 cores, it requires $\sim$ 22 hours of time
\footnote{A machine with Intel(R) Core(TM) i7-10700 CPU 
16 GB RAM is used for all the computations.}.

\begin{figure}[!htb]
\includegraphics[width=0.24\textwidth]{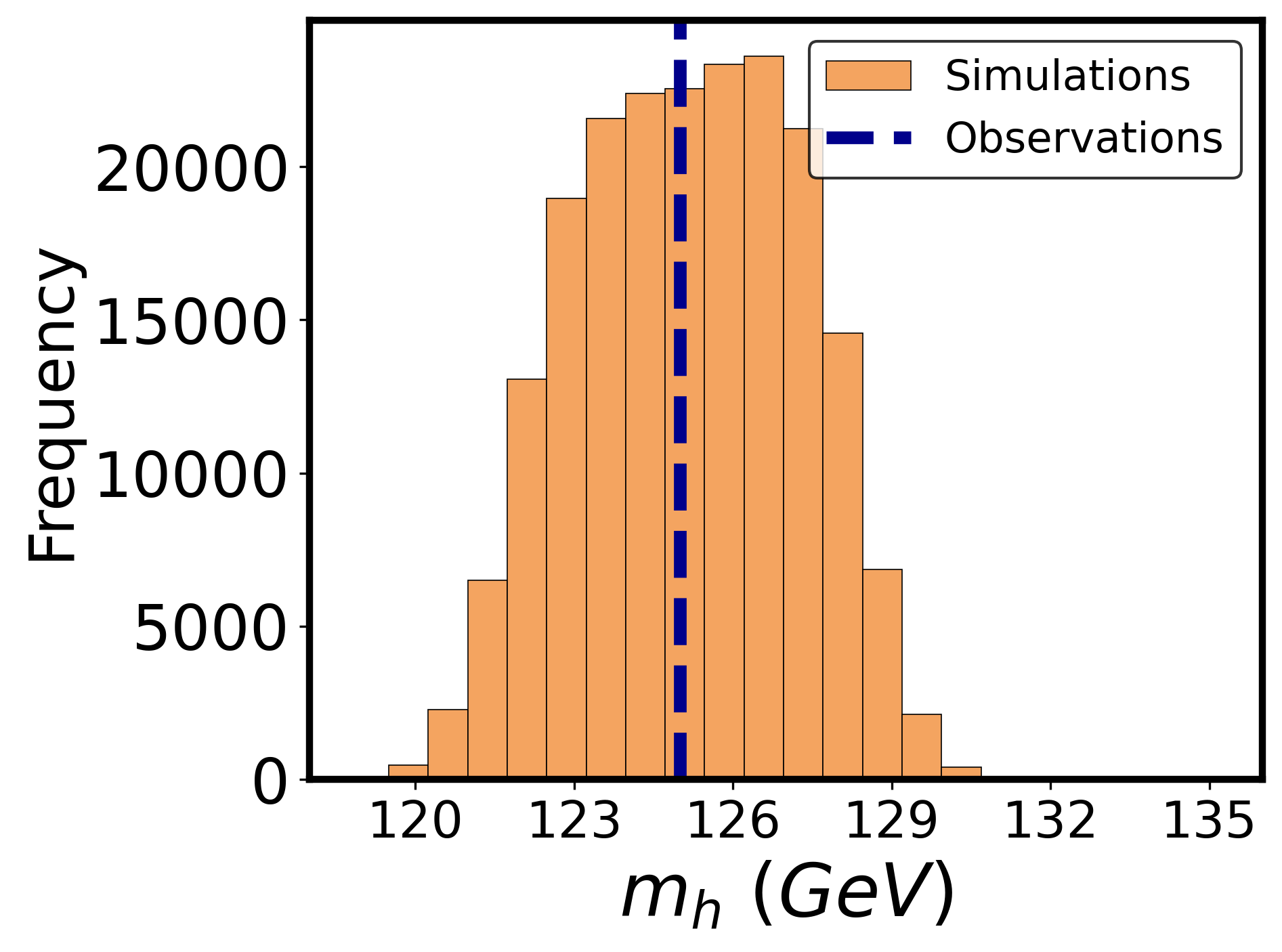}
\includegraphics[width=0.24\textwidth]{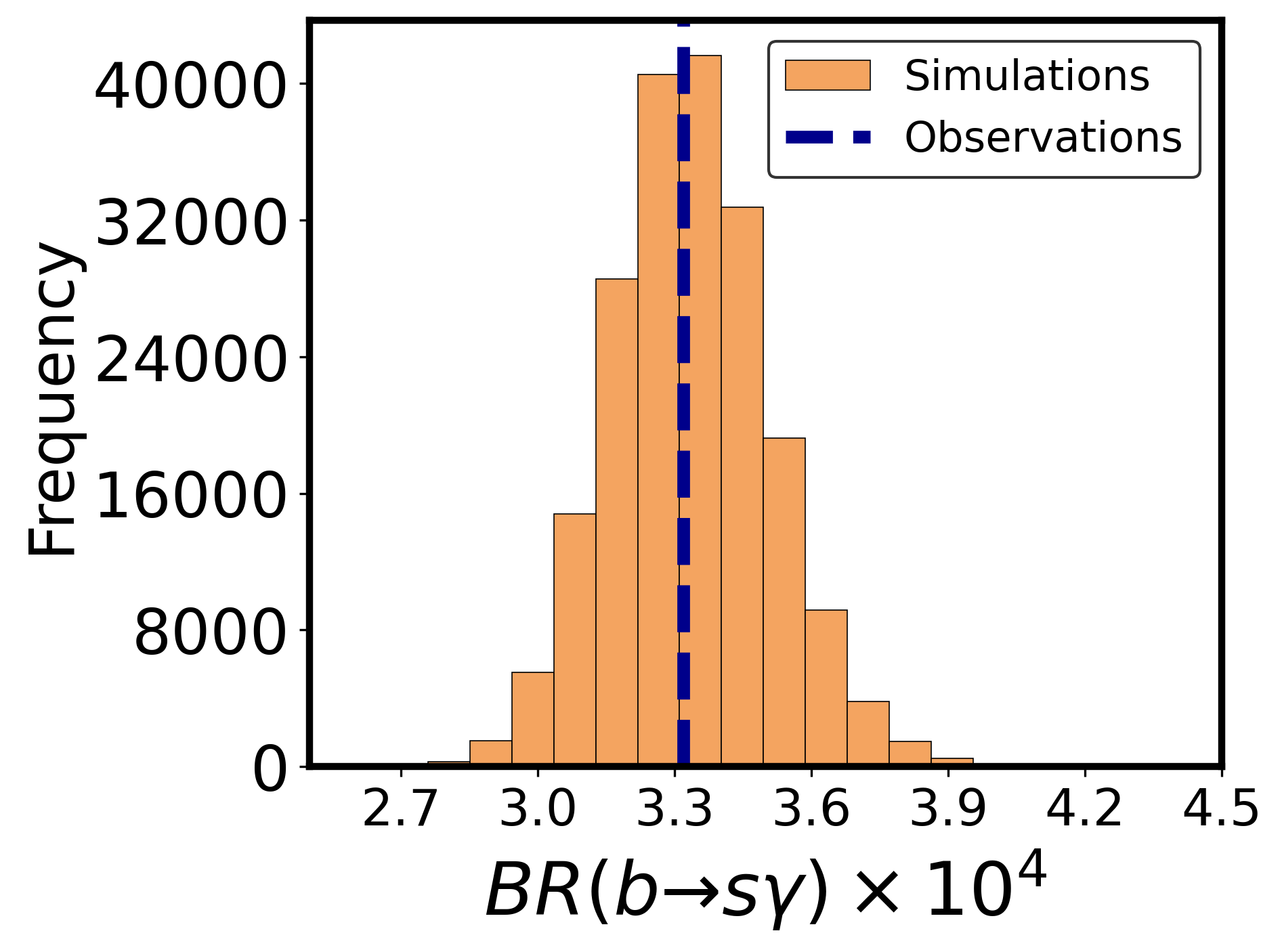}
\includegraphics[width=0.24\textwidth]{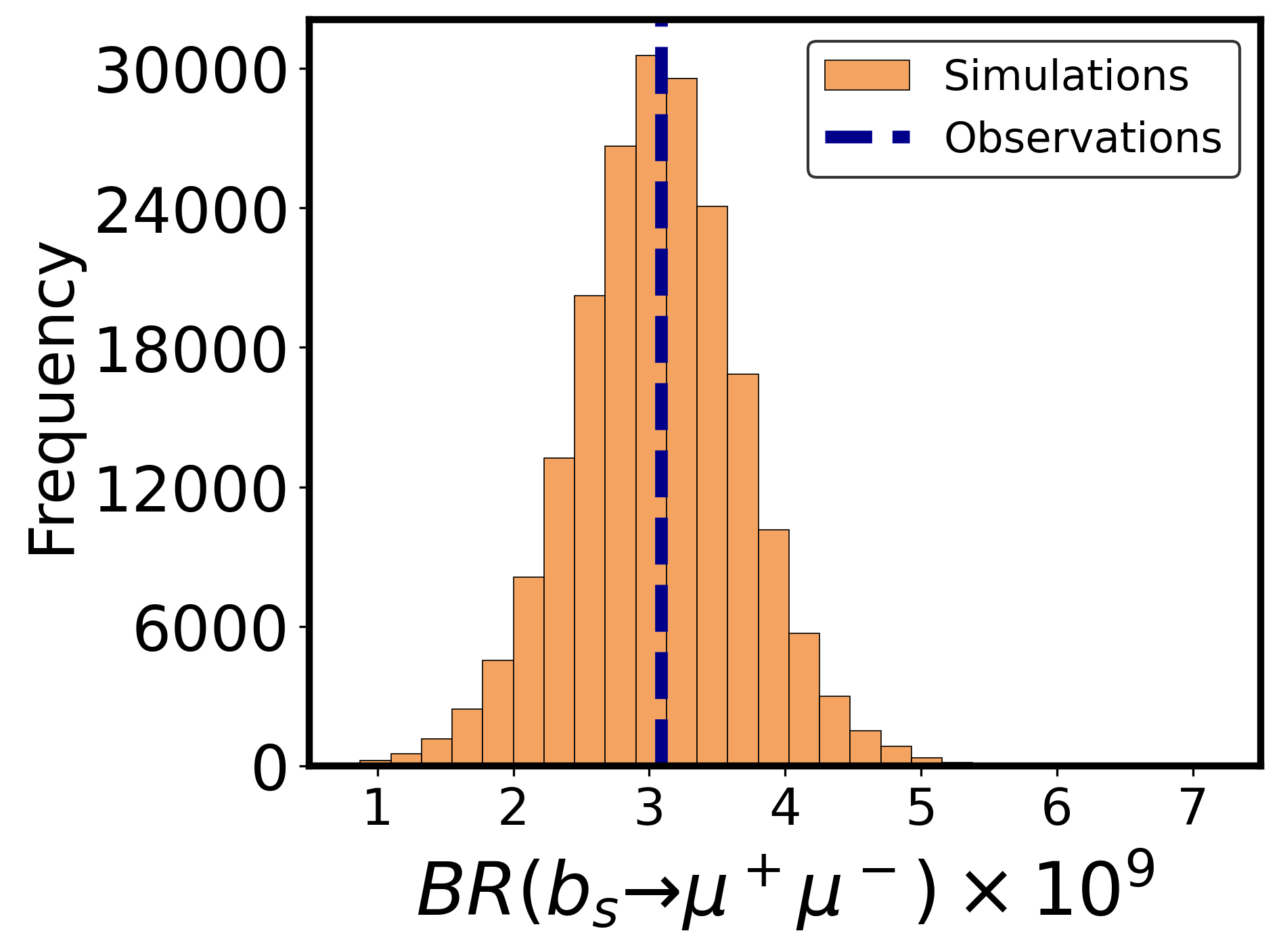}
\includegraphics[width=0.24\textwidth]{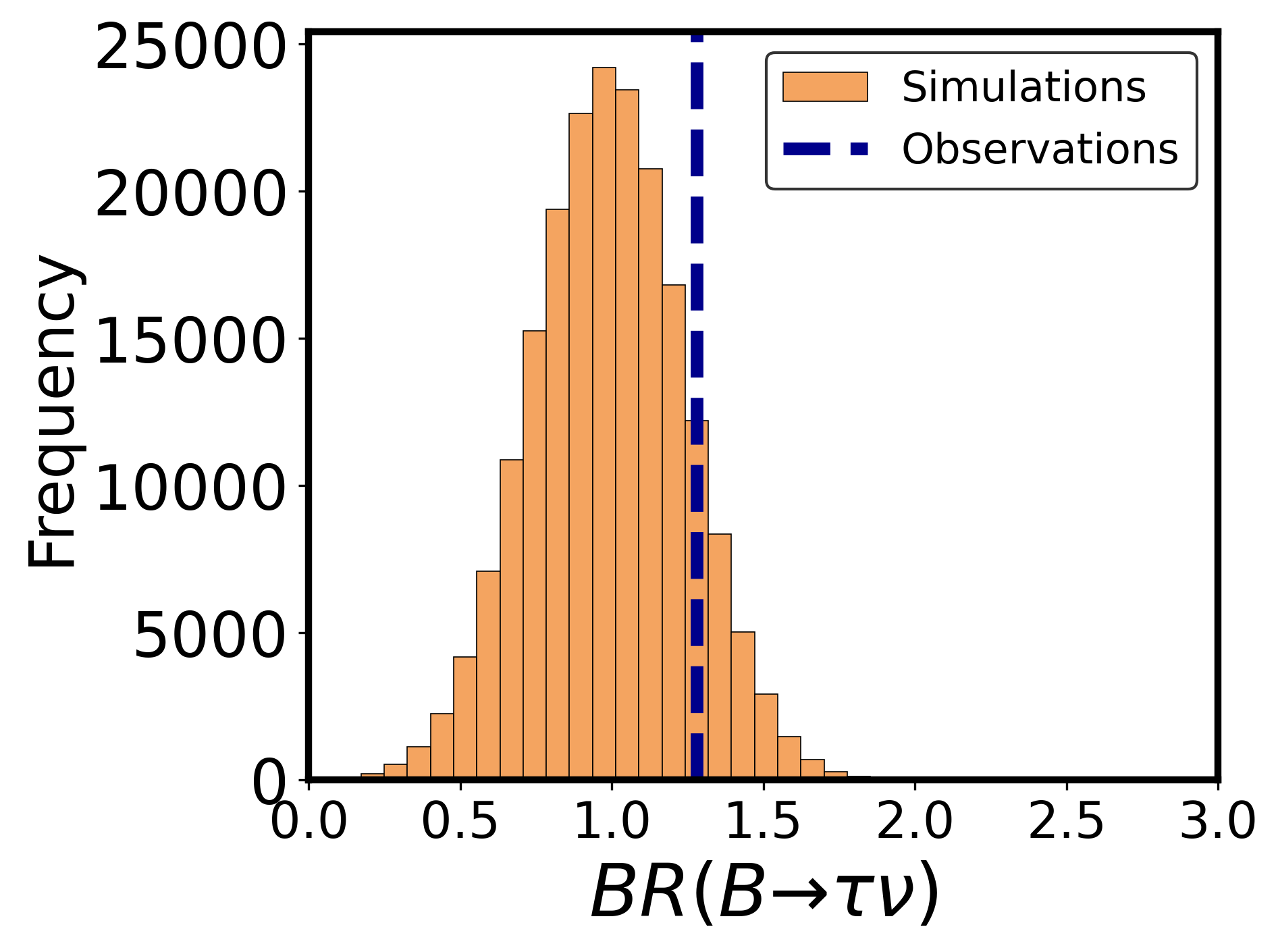}\\
\includegraphics[width=0.19\textwidth]{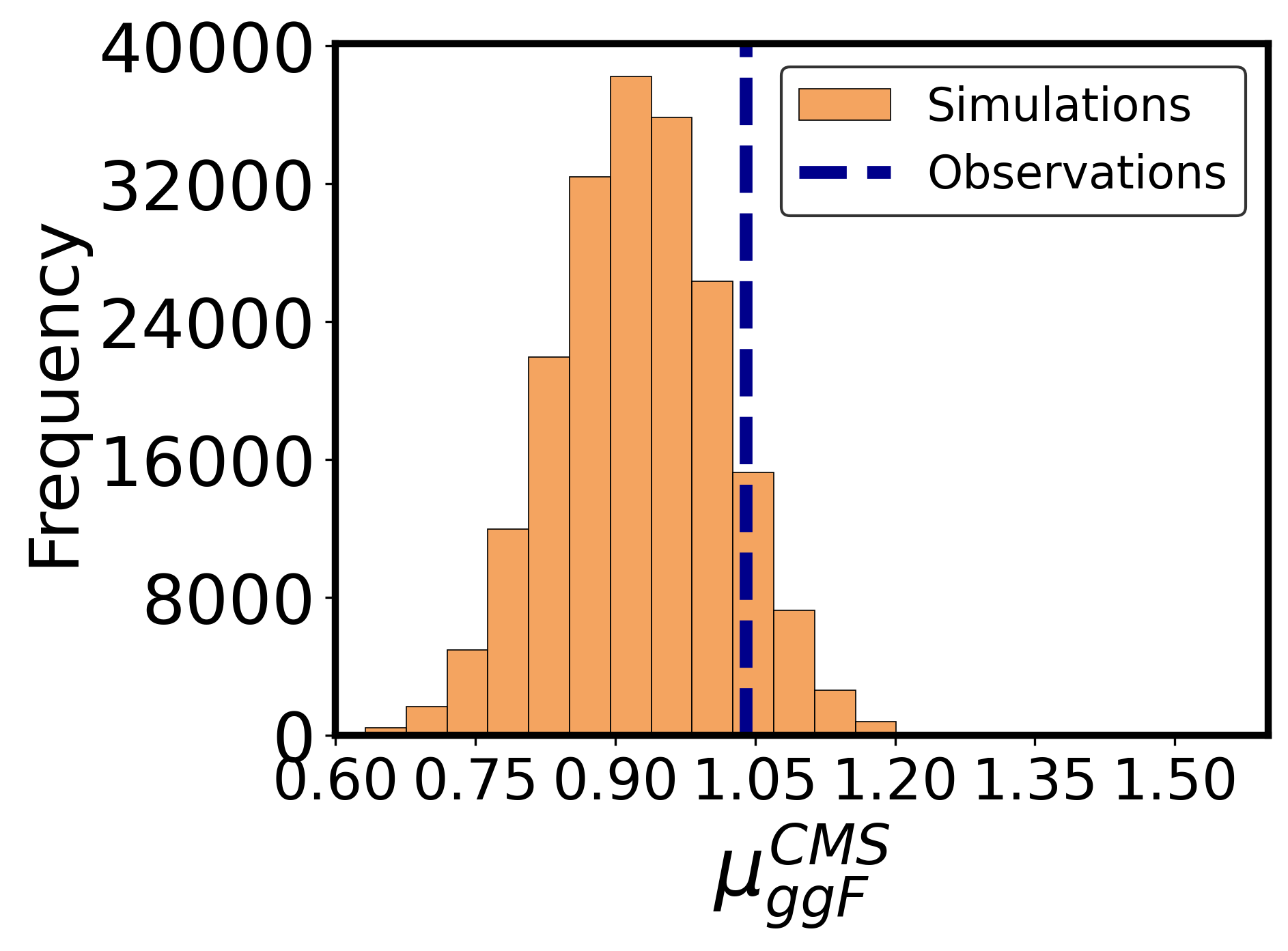}
\includegraphics[width=0.19\textwidth]{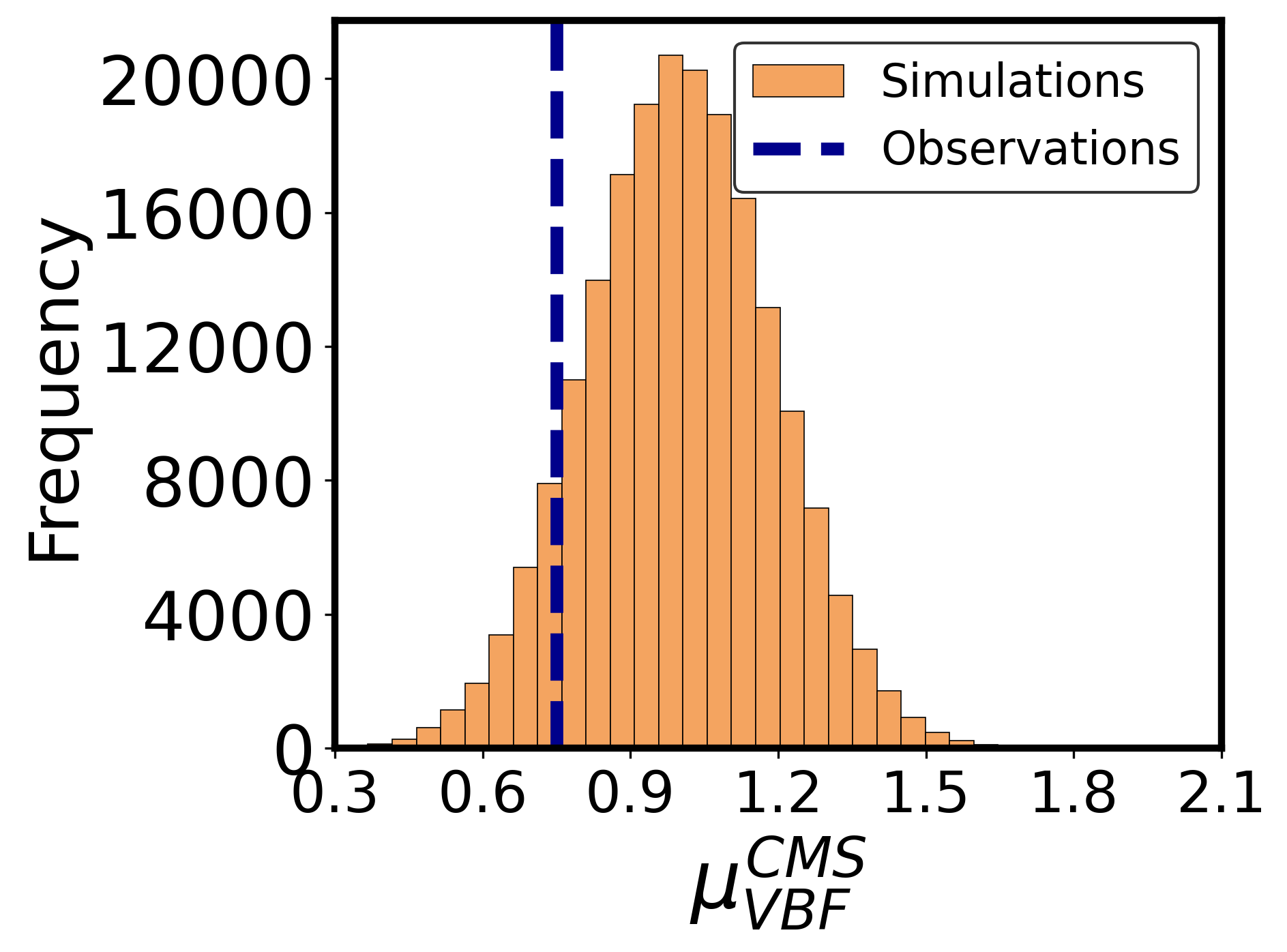}
\includegraphics[width=0.19\textwidth]{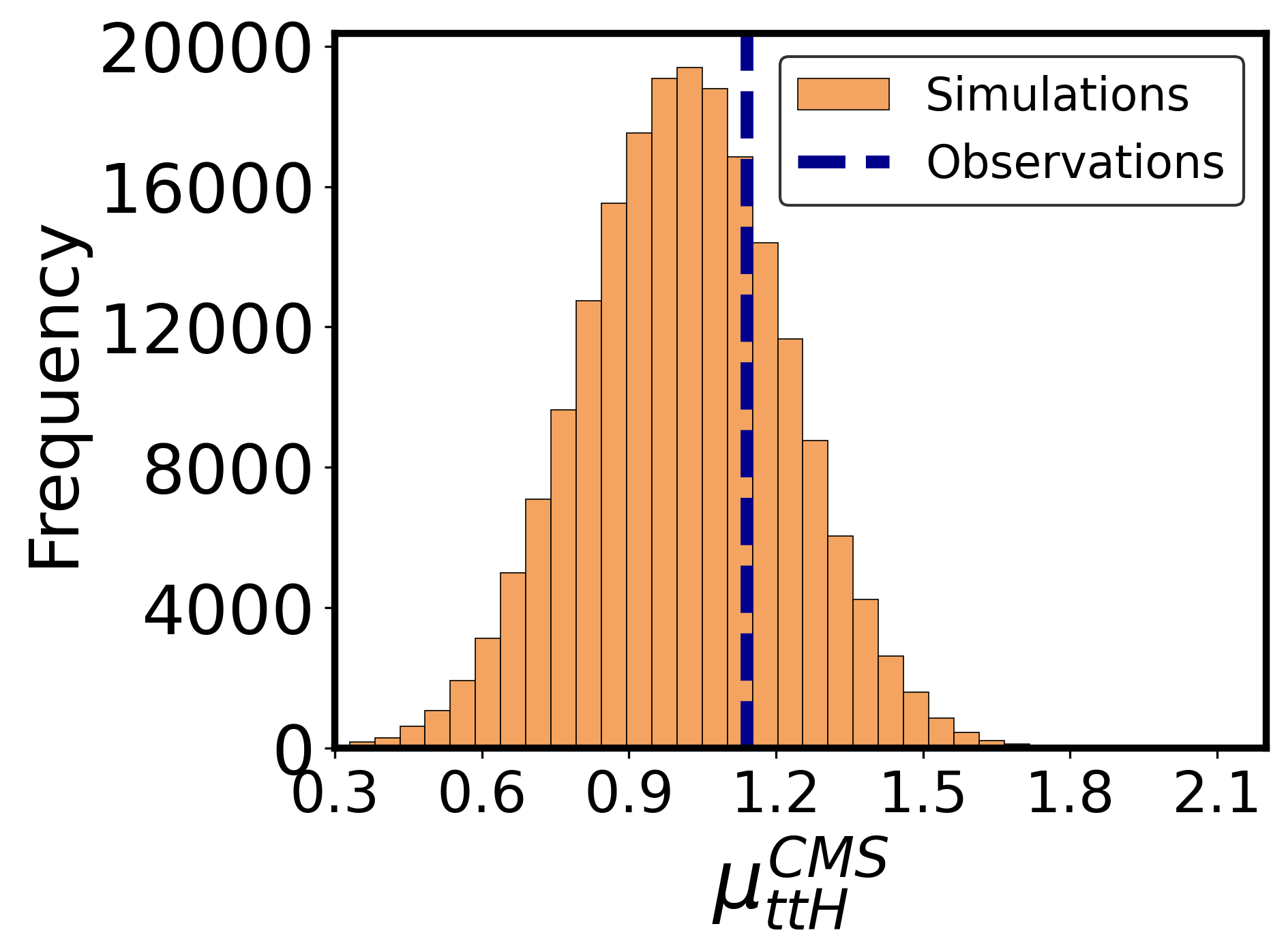}
\includegraphics[width=0.19\textwidth]{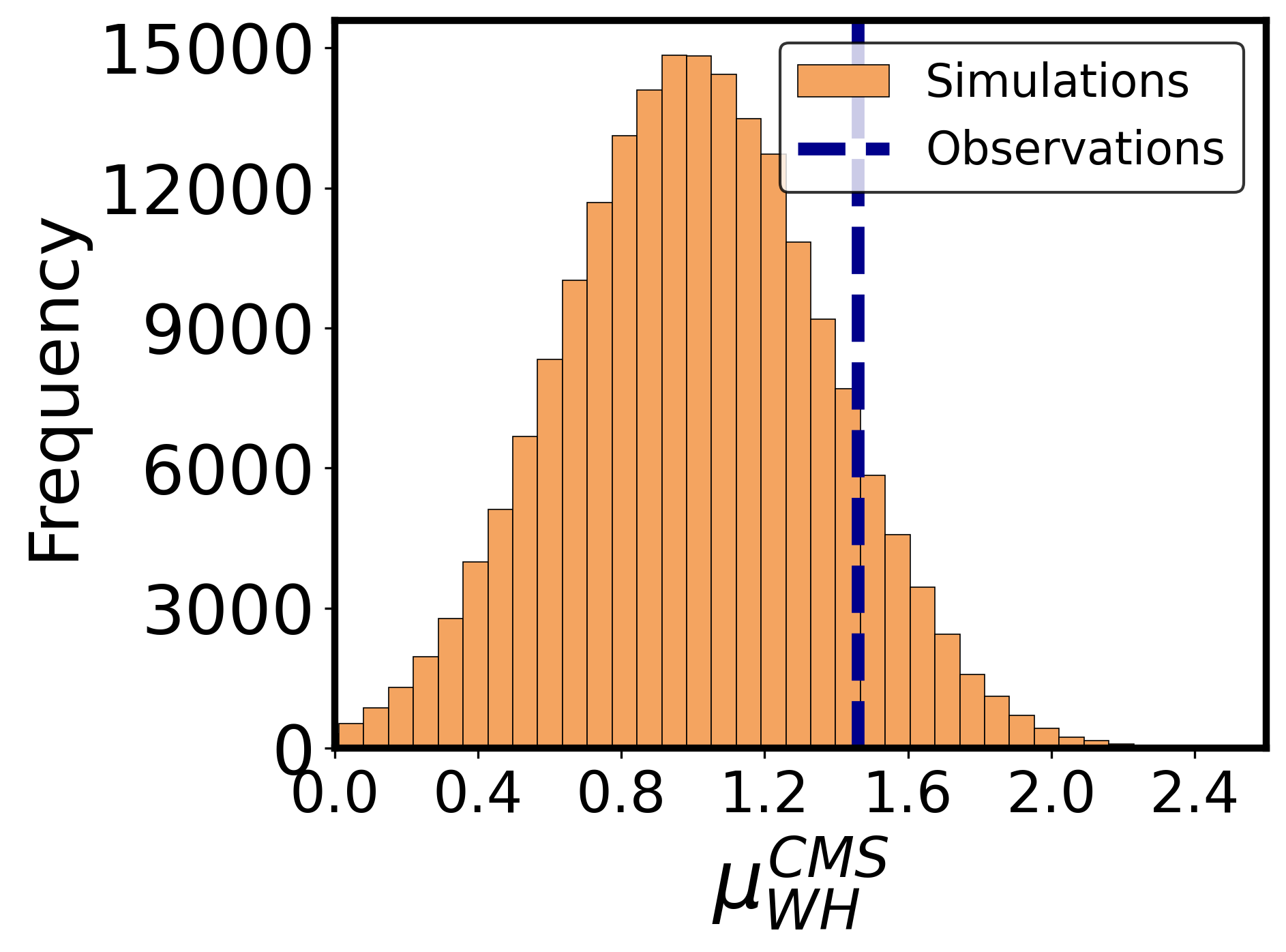}
\includegraphics[width=0.19\textwidth]{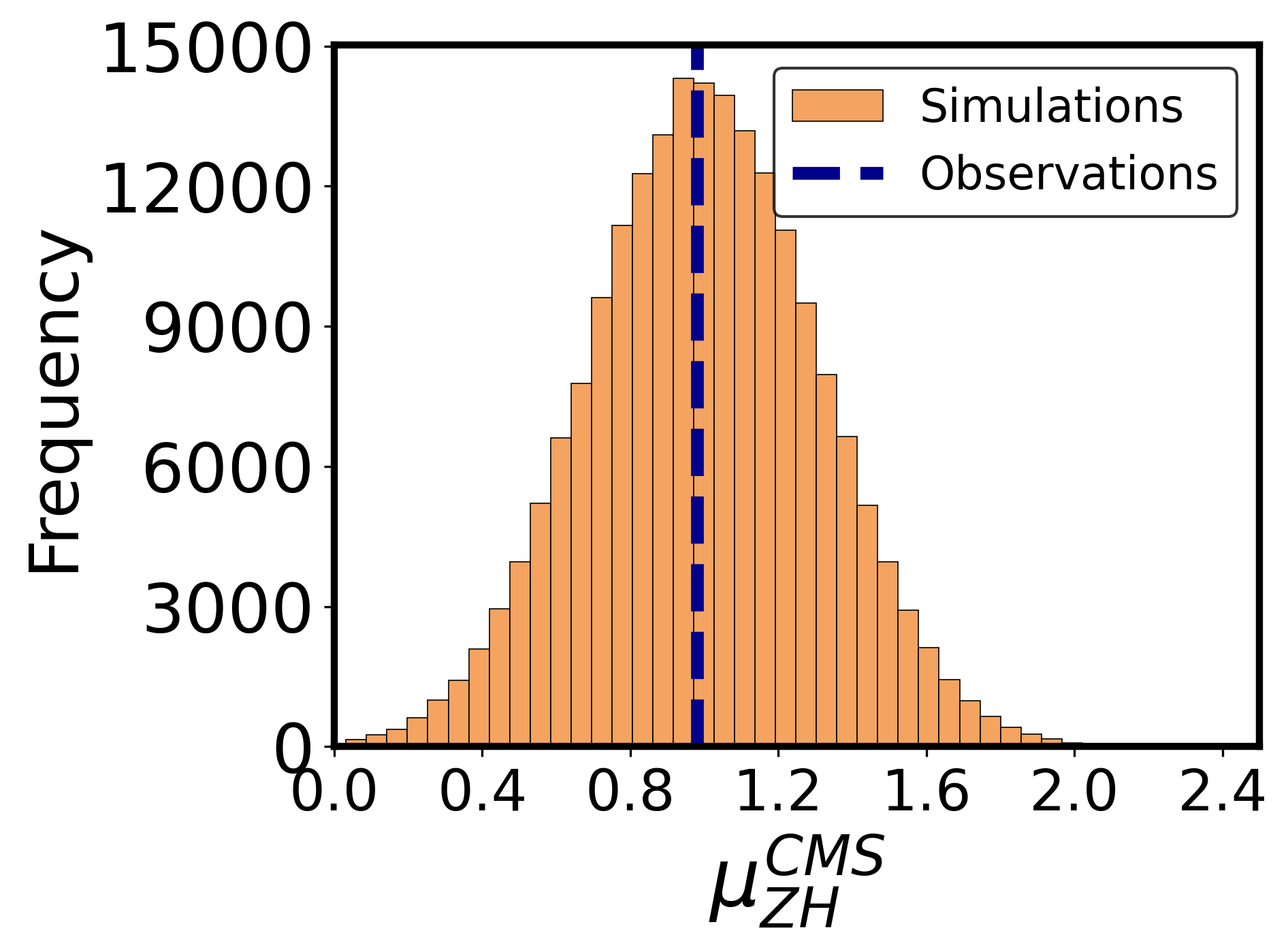}\\
\includegraphics[width=0.19\textwidth]{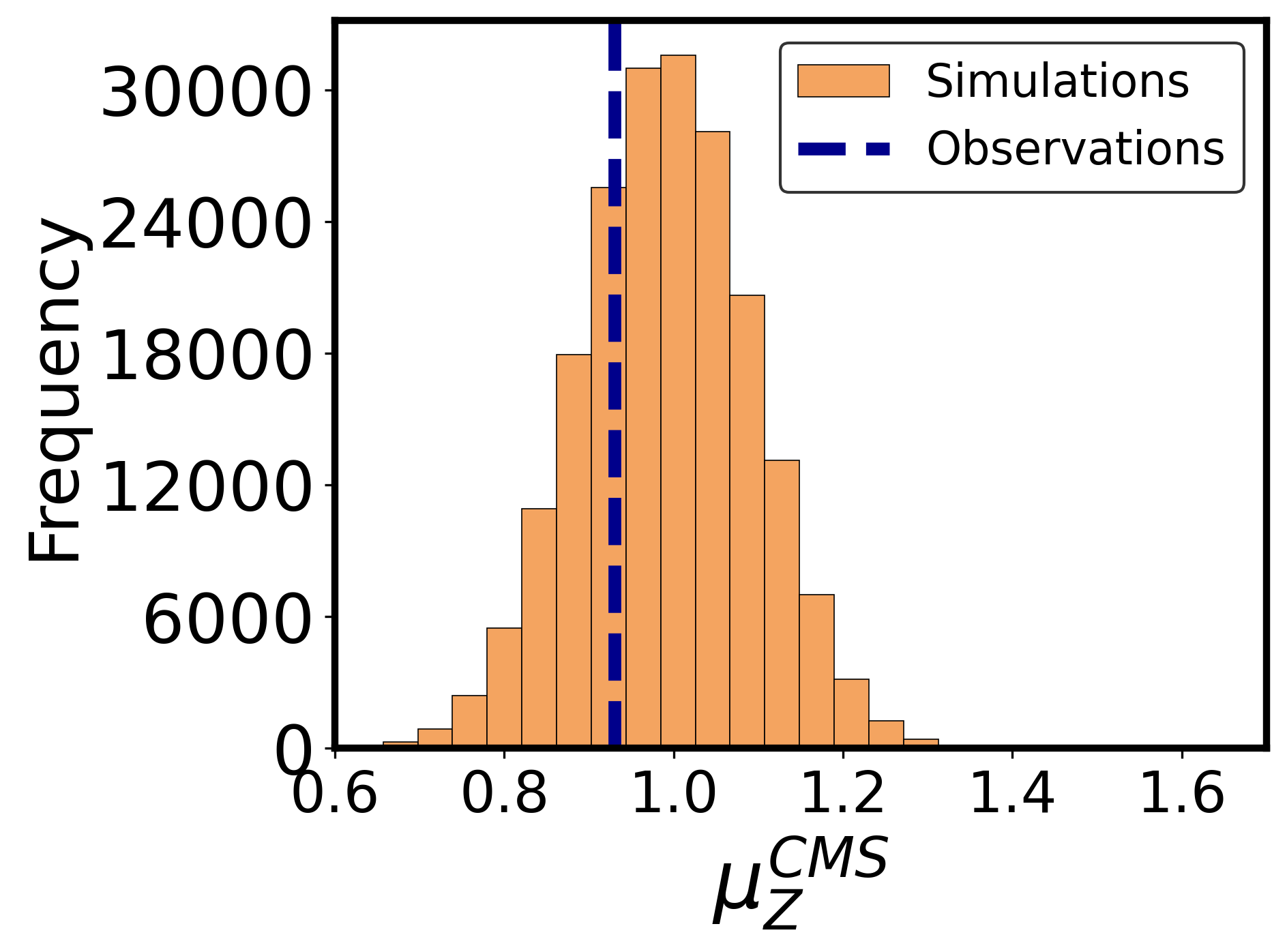}
\includegraphics[width=0.19\textwidth]{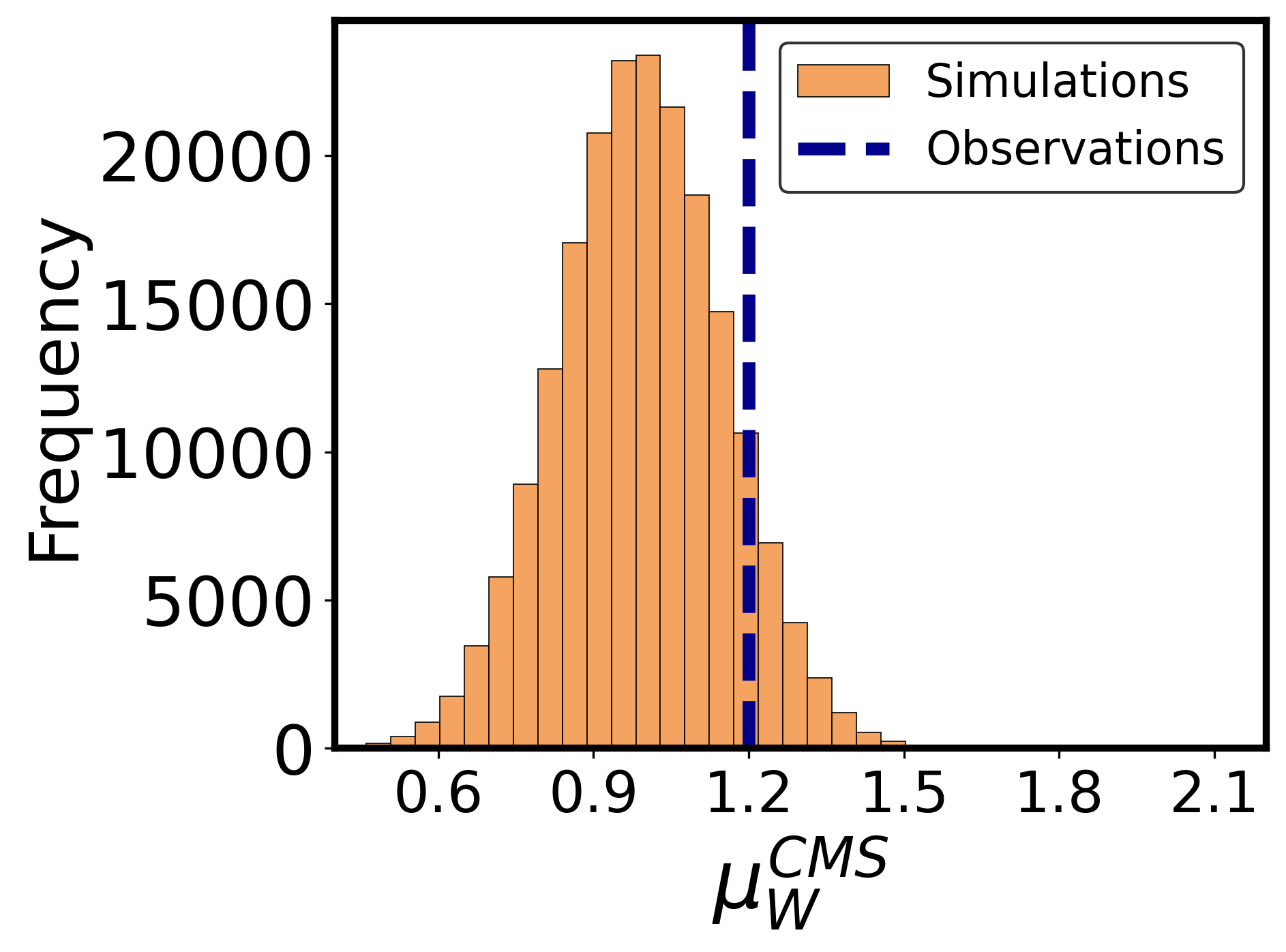}
\includegraphics[width=0.19\textwidth]{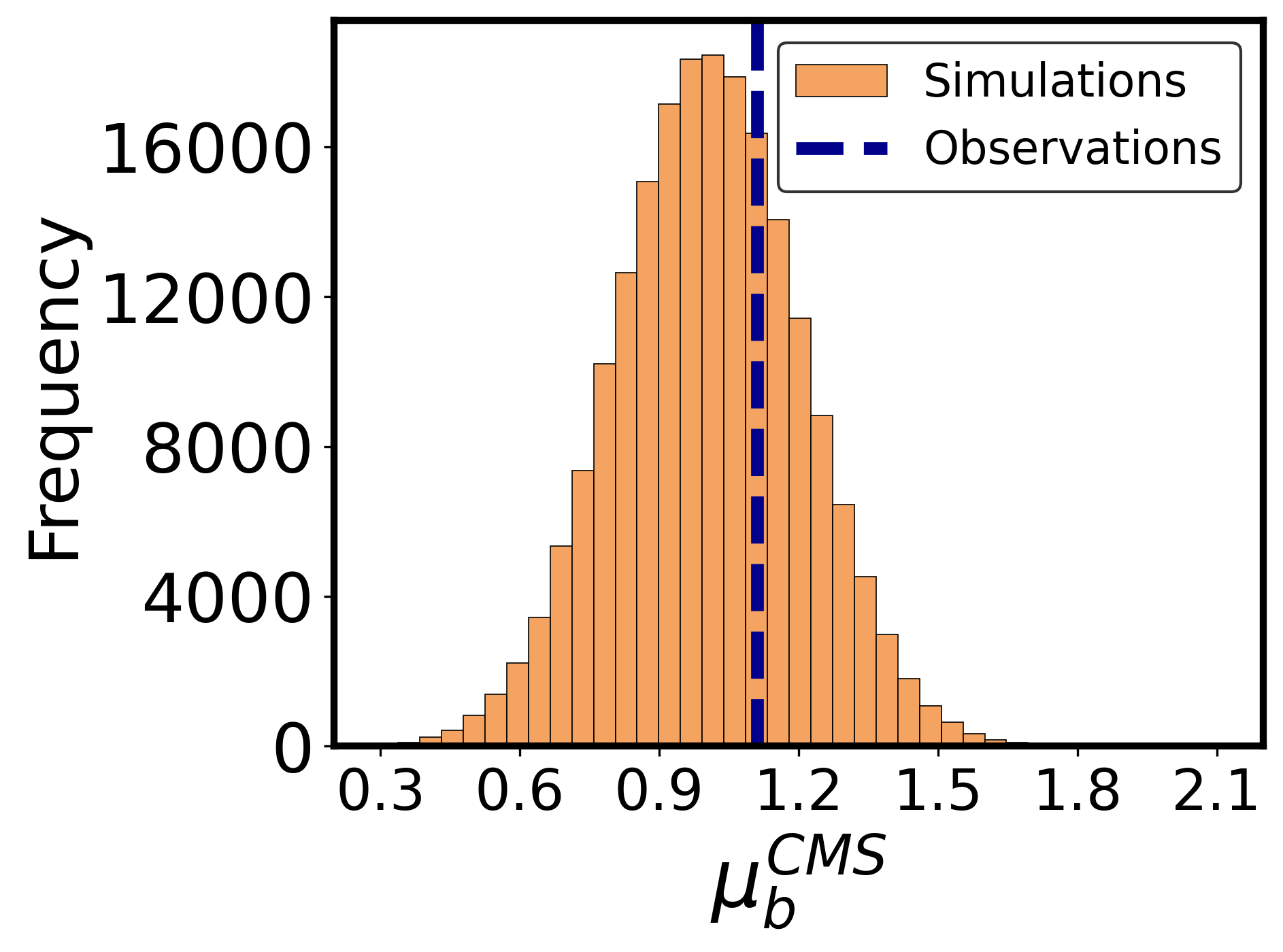}
\includegraphics[width=0.19\textwidth]{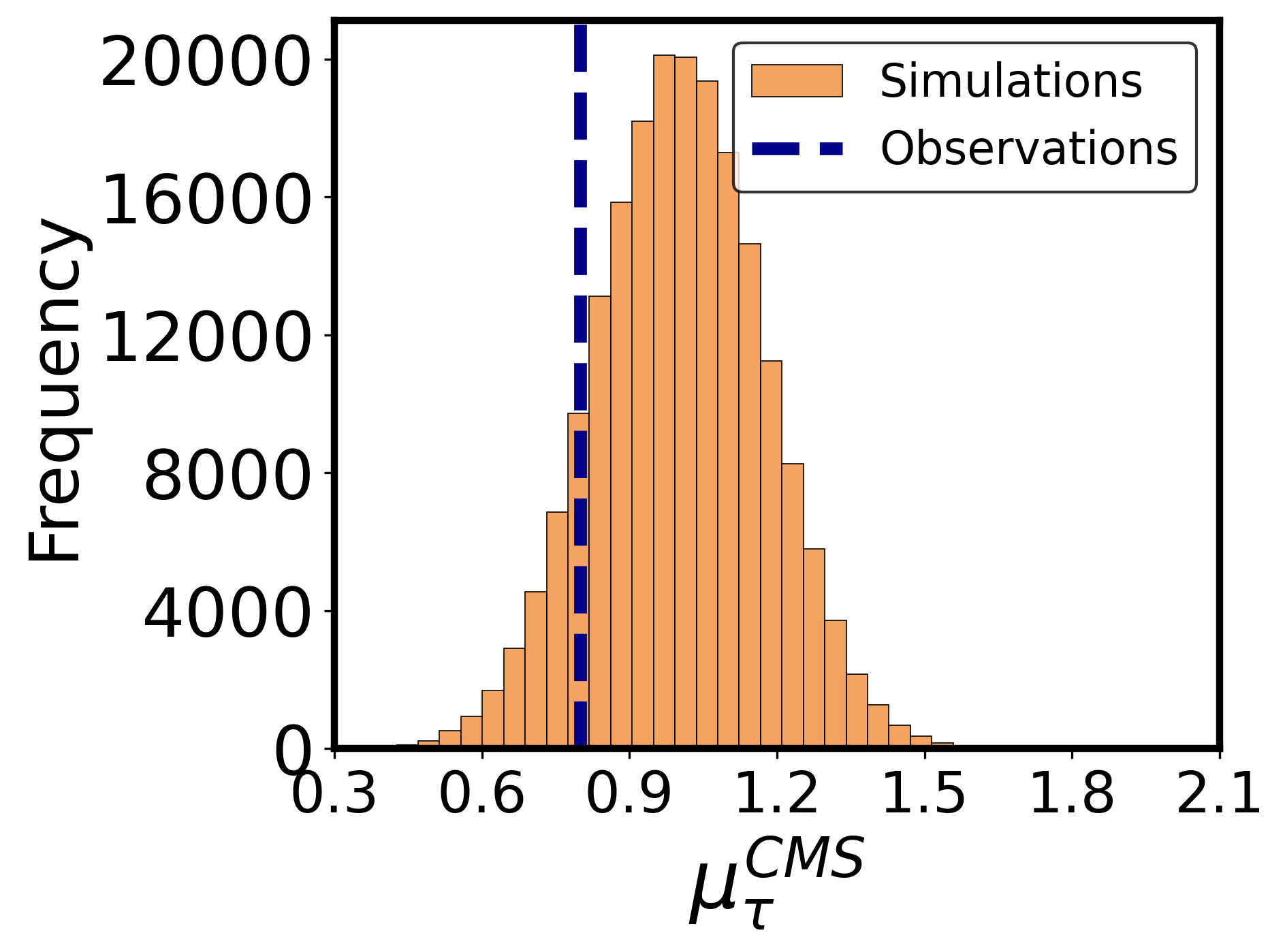}
\includegraphics[width=0.19\textwidth]{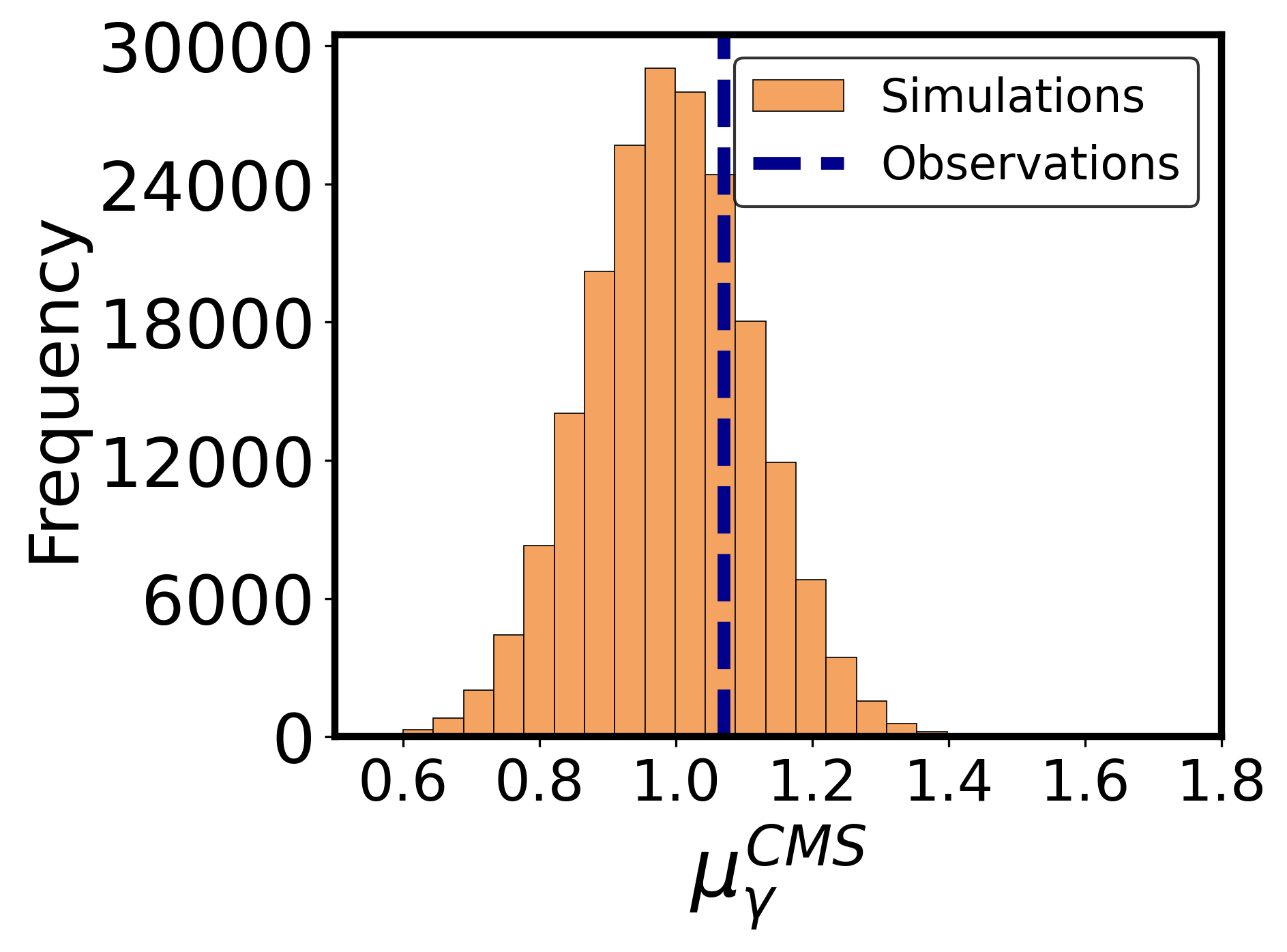}
\caption{Distributions of all the 14 observables in training sample after adding the noise 
are displayed. The orange-brown histograms represent the simulated training sample distribution whereas the dashed blue lines indicate the true value of each observable.}
    \label{fig:obs_dist}
\end{figure}


To make the observables, calculated by \texttt{FeynHiggs}, resemble the 
experimental data, we need to add some noise or error to the observables  in the form $\mathcal{N}(0, \sigma^2)$, where the $\sigma$ corresponding to each observable is already mentioned in the Table.~\ref{tab:observable}. So, the observable, $f(x)$, generated by the packages will be updated to  $f^{\prime}(x) = f(x) + \mathcal{N}(0, \sigma^2)$. The training sample must be centered around the true values, as required by the SBI methods~\cite{sbilink}. To verify this, we plot the histogram of each observable after adding noise, as shown in Figure~\ref{fig:obs_dist}. The plot indicates that not all distributions are centered precisely around the true values. However, it is noteworthy that the true values, represented by the dashed dark blue lines, lie within the distribution of simulated observables, depicted by the light orange-brown histogram. Parameters often have different scales and ranges, which can bias the model or lead to numerical instability, such as overflow or underflow during computation. Normalizing parameters to the range [0, 1] ensures a stable and balanced learning process. Once the posterior samples are drawn, we denormalize them to retrieve the parameters in their original scale for meaningful interpretation.


We first determine the minimum number of samples required to accurately reconstruct the posterior of the model parameters. For that, we gradually increase  the number of samples ($1\times 10^3$, $2\times 10^3$, $5\times 10^3$ ...) in the training dataset from our previously stored $2\times 10^5$ samples (as mentioned in Section~\ref{sec:data_file}). For each case, 	95\% of the samples are allocated for training, and 5\% are reserved for validation of the SBI model using the TARP test. After training, posterior samples are drawn around the central values of each observable. To measure the posterior sample efficiency, we input the posterior samples into the \texttt{FeynHiggs} package to calculate the corresponding observables. For each parameter set, we check whether the observables lie within $3\sigma$ of the values specified in Table~\ref{tab:observable}. The posterior sample efficiency is defined as the ratio of parameter sets meeting the $3\sigma$ condition to the total number of posterior samples generated. We also compare the time consumption of each algorithm corresponding to different sample subsets. 


\subsection{Ablation study and hyperparameter set-up}
\label{sec:hyper_param}

Since each SBI method uses a neural network with distinct architectures and features, we now carry out an ablation study to find out the best set of hyperparameters. 
  In both NPE and NLE methods, we use \texttt{MAF} ``\texttt{density\_estimator}'', whereas for NRE, we use \texttt{ResNET},  as already mentioned in Sections ~\ref{sec:density_function} and ~\ref{sec:nre}.
  There are different hyperparameters for \texttt{MAF} density function such as ``\texttt{hidden\_features}'' (\texttt{HF}) and ``\texttt{num\_transforms}'' (\texttt{NT}) along with other ML hyperparameters like ``\texttt{Training Batch size}'' (\texttt{TBS}), ``\texttt{Learning rate}'' (\texttt{LR}) etc.
\begin{table}[!htb]
\resizebox{\columnwidth}{!}{%
\centering
\begin{tabular}{||c|c|c|c|c|c|c|c|c|c||}
\hline\hline
Fixed Part & & \multicolumn{8}{c||}{Efficiency with varying hyperparameter} \\
\hline
\texttt{MAF} 1 layer & \multirow{2}{*}{\texttt{HF} value} & \multirow{2}{*}{2}& \multirow{2}{*}{10} &\multirow{2}{*}{50} & \multirow{2}{*}{100} & \multirow{2}{*}{400} &\multirow{2}{*}{600}& \multirow{2}{*}{800} & \multirow{2}{*}{1000} \\
\texttt{NT}=20, \texttt{TBS}=512 &  & & & & & & & & \\
\cline{2-10}
LR=$0.0001$ & Efficiency (\%) & 96.67 & 97.29 & 98.70 & 99.34 & 99.51 & 98.88 & 97.68 & 96.88 \\
\hline\hline
\texttt{MAF} 1 layer & \multirow{2}{*}{\texttt{NT} value} & \multirow{2}{*}{1}& \multirow{2}{*}{2} &\multirow{2}{*}{5} & \multirow{2}{*}{10} & \multirow{2}{*}{20} &\multirow{2}{*}{40}& \multirow{2}{*}{70} & \multirow{2}{*}{100} \\
\texttt{HF}=400, \texttt{TBS}=512 &  & & & & & & & & \\
\cline{2-10}
LR=$0.0001$ & Efficiency (\%) & 96.4 & 97.10 & 97.80 & 98.88 & 99.51 & 99.56 & 99.60 & 99.62 \\
\hline\hline
\texttt{MAF} 1 layer & \multirow{2}{*}{\texttt{TBS} value} & \multirow{2}{*}{16}& \multirow{2}{*}{32} &\multirow{2}{*}{64} & \multirow{2}{*}{128} & \multirow{2}{*}{256} &\multirow{2}{*}{512}& \multirow{2}{*}{1024} & \multirow{2}{*}{2048} \\
 \texttt{HF}=400, \texttt{NT}=20 &  & & & & & & & & \\
\cline{2-10}
LR=$0.0001$ & Efficiency (\%) & 98.09 & 98.87 & 99.16 & 99.27 & 99.45 & 99.51 & 98.96 & 97.60 \\
\hline\hline
\texttt{MAF} 1 layer & \multirow{2}{*}{\texttt{LR} value} & \multirow{2}{*}{0.00001}& \multirow{2}{*}{0.0001} &\multirow{2}{*}{0.001} & \multirow{2}{*}{0.01} & \multirow{2}{*}{0.1} &\multirow{2}{*}{0.3}& \multirow{2}{*}{0.5} & \multirow{2}{*}{1.0} \\
\texttt{HF}=400, \texttt{NT}=20 &  & & & & & & & & \\
\cline{2-10}
\texttt{TBS}=512 & Efficiency (\%) & 98.68 & 99.51 & 99.56 & 99.27 & 99.05 & 98.78 & 98.06 & 97.60 \\
\hline\hline
\texttt{HF}=400, \texttt{NT}=20 & \texttt{MAF} layer & \multicolumn{2}{c|}{1} & \multicolumn{2}{c|}{2} & \multicolumn{2}{c|}{3} & \multicolumn{2}{c|}{4}\\
\cline{2-10}
\texttt{TBS}=512, \texttt{LR}=0.0001 & Efficiency (\%) & \multicolumn{2}{c|}{99.51} & \multicolumn{2}{c|}{99.32} & \multicolumn{2}{c|}{99.02} & \multicolumn{2}{c|}{99.60} \\
\hline\hline
\end{tabular}
} 
\caption{The effect of  hyperparameter variation  on posterior sample efficiency is illustrated for the NPE algorithm with $1\times 10^5$ samples.}
\label{tab:param_opt}
\end{table}
In \texttt{LtU-ILI}, one can define \texttt{nets}  by using multiple layers of the same or different density estimators/functions, which have their own features that 
can be adjusted to analyze their impact on the posterior sample efficiency.   
  Ideally, one should perform the hyperparameter optimization by varying all the elements of the networks simultaneously. However, due to limited computational resources, we restrict ourselves to an ablation study where we only vary certain parts of the network while keeping the rest at some fixed values. The ablation study results for NPE algorithm are presented in Table.~\ref{tab:param_opt}.

From the ablation study in Table~\ref{tab:param_opt}, it is evident that the efficiency decreases both for very low and high values of the \texttt{HF} parameter. The efficiency reaches its maximum around \texttt{HF} = 400 when other parameters are fixed at the values specified in the table. A similar trend is observed for the \texttt{TBS} parameter. On the other hand, the efficiency consistently increases with larger \texttt{NT} values. However, it is worth pointing out that the improvement saturates after a certain threshold (\texttt{NT} = 20). 
We conducted a similar study for the NLE and NRE algorithms. However, as shown in Section~\ref{sec:posterior_eff}, NPE yields the more promising results compared to others. Therefore, we have not discussed their ablation study in detail. The 
optimized hyperparameters considered for each algorithm, along with their values, are listed in Table~\ref{tab:hyper_param}.
   
   \begin{table}[!htb]
   	\resizebox{\columnwidth}{!}{%
   		\centering
   		\begin{tabular}{||c|c|c|c||}
   			\hline\hline
   			Features & NPE & NLE & NRE \\
   			\hline\hline
   			Density function & \multirow{2}{*}{\texttt{MAF}} & \multirow{2}{*}{\texttt{MAF}} & \multirow{2}{*}{\texttt{ResNET}} \\ 
   			used (1 layer)           &                               &                               &                                 \\ 
   			\hline
   			Density function  &  ``\texttt{hidden\_features}''=400  & ``\texttt{hidden\_features}''=200 & ``\texttt{hidden\_features}''=200 \\
   			features & ``\texttt{num\_transforms}''=20 & ``\texttt{num\_transforms}''=20 & ``\texttt{num\_blocks}''=20 \\
   			\hline\hline
   			\multicolumn{4}{||c||}{Common hyperparameters} \\
   			\hline
   			\multicolumn{4}{||c||}{\texttt{Batch size} = 512, \texttt{Learning rate} = $10^{-4}$ }\\
   			\hline\hline
   		\end{tabular}
   	} 
   	\caption{The density function along with the details of optimized  hyperparameter are shown here for each algorithm}
   	\label{tab:hyper_param}
   \end{table}


\subsection{TARP test results} 
\label{sec:result_tarp}

\begin{figure}[!htb]
\centering
\includegraphics[width=0.3\textwidth]{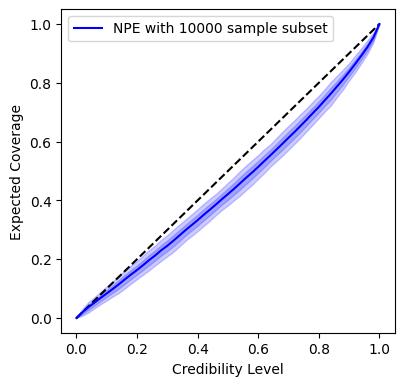}
\includegraphics[width=0.3\textwidth]{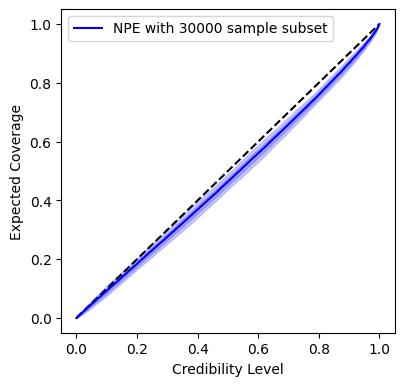}
\includegraphics[width=0.3\textwidth]{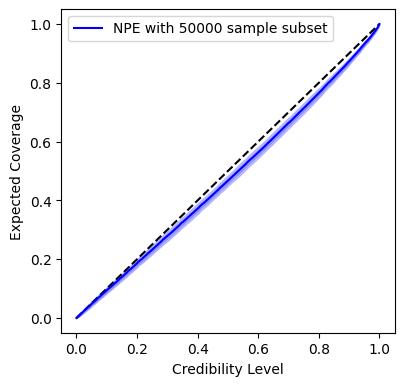}\\
\includegraphics[width=0.3\textwidth]{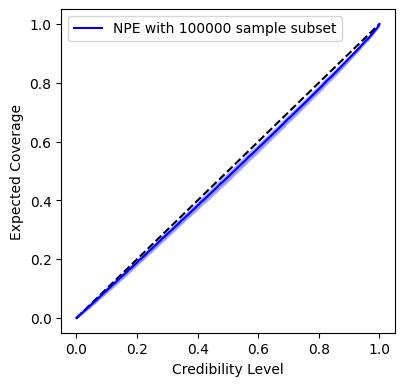}
\includegraphics[width=0.3\textwidth]{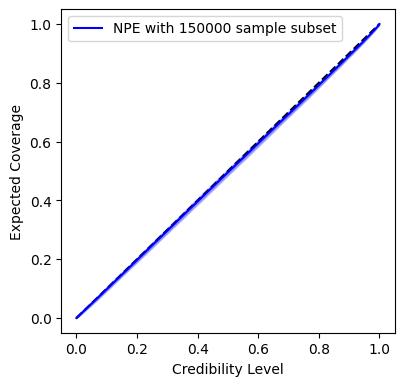}
\includegraphics[width=0.3\textwidth]{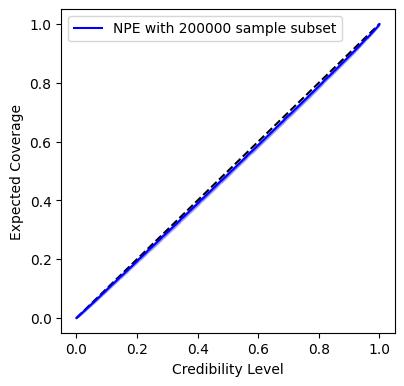}
\caption{TARP test plot corresponding to $1\times 10^4$, $3\times10^4$, $5\times10^4$, $1\times 10^5$, $1.5\times10^5$, $2\times10^5$ sample subsets with NPE method are displayed here. }
    \label{fig:tarp_npe}
\end{figure}

In this section, we discuss the TARP test results in detail. 
Figure~\ref{fig:tarp_npe} illustrates the TARP plot for various subsets corresponding to the NPE algorithm. It is important to note that TARP test results are presented only for subsets with more than $5\times 10^3$ samples, as smaller subsets tend to yield less meaningful or unreliable outcomes. As mentioned in \cite{2023PMLR..20219256L} if the blue line (the computed ECP) follows the dashed black curve, closely, it signifies that the density estimator is unbiased (ideal case). 
If uncertainties are biased (either under or over confident) then the blue curve will deviate from the black dashed curve giving rise to a ``S'' shaped curve. One of the ways to quantitatively describe this deviation is to compute the deviation at different credibility levels as shown in Table~\ref{tab:tarp_diff}. 

   
\begin{table}[!htb]
   		\centering
   		\begin{tabular}{||c|c||c|c|c||}
   			\hline\hline
   		    Method & Sample & $\Delta y_{25}$ & $\Delta y_{50}$ & $\Delta y_{75}$ \\
   			\hline\hline
   			\multirow{6}{*}{NPE} & $1\times 10^4$ & 0.03 & 0.08 & 0.07 \\ \cline{2-5}
   			 & $3\times 10^4$ & 0.02 & 0.03 & 0.04 \\ \cline{2-5}
   			 & $5\times 10^4$ & 0.015 & 0.04 & 0.04 \\ \cline{2-5}
   			 & $1\times 10^5$ & 0.011 & 0.012 & 0.016 \\ \cline{2-5}
   			 & $1.5\times 10^5$ & 0.002 & 0.01 & 0.003 \\ \cline{2-5}
   			 & $2\times 10^5$ & 0.016 & 0.02 & 0.01 \\
   			 \hline
   			 NLE & $2\times 10^5$ & 0.09 & 0.34 & 0.59 \\
   			 \hline
   			 NRE & $2\times 10^5$ & 0.09 & 0.34 & 0.59 \\ 
   			\hline\hline
   		\end{tabular}
   	\caption{Deviation of expected coverage of TARP plot from the accurate posterior line at three different C.L. points are shown for all the sample subsets in three SBI methods.}
   	\label{tab:tarp_diff}
   \end{table}

\begin{figure}[!htb]
\centering
\includegraphics[width=0.49\textwidth]{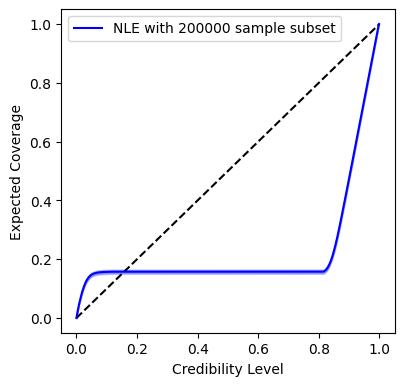}
\includegraphics[width=0.49\textwidth]{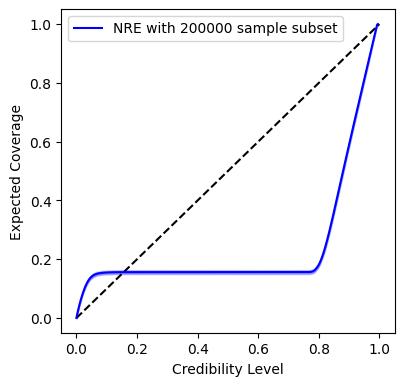}
\caption{TARP test plot corresponding to $2\times 10^5$ sample subset for NLE (Left) and  NRE (right) method are displayed here. }
    \label{fig:tarp_com}
\end{figure}
From Figure~\ref{fig:tarp_npe}, it is clear that there is no significant improvement in the TARP plot beyond the $1\times 10^5$ sample subset for our model. The difference in the expected coverage from the accurate posterior line (dashed straight line) is also shown in Table~\ref{tab:tarp_diff}. 
This difference, denoted as $\Delta y$, is calculated at three different credibility levels (C.L.): $\Delta y_{25}$, $\Delta y_{50}$, and $\Delta y_{75}$, corresponding to 0.25, 0.50, and 0.75 C.L., respectively. It is evident that the TARP method with $1.5 \times 10^5$ samples provides the best results. However, there is no significant improvement observed when increasing the sample subset beyond $1\times 10^5$. 
Additionally, Figure~\ref{fig:tarp_com} presents the TARP test results for the other two algorithms, focusing solely on the $2\times 10^5$ sample set. As is evident from the figure, neither of these two algorithms performs satisfactorily in the TARP test. We have presented the $\Delta y$ values for these two methods in Table~\ref{tab:tarp_diff}, and both show large deviations from the accurate posterior line. It is also noteworthy that similar outcomes were observed for other sample subsets with these algorithms.
Based on the TARP test results, we conclude that the NPE algorithm demonstrates superior performance for the pMSSM model under consideration. This trend will be further corroborated by additional results, including posterior sample efficiency and time consumption.

\subsection{Posterior sample efficiency and time consumption}
\label{sec:posterior_eff}
\begin{figure}[!htb]
\centering
\includegraphics[width=0.49\textwidth]{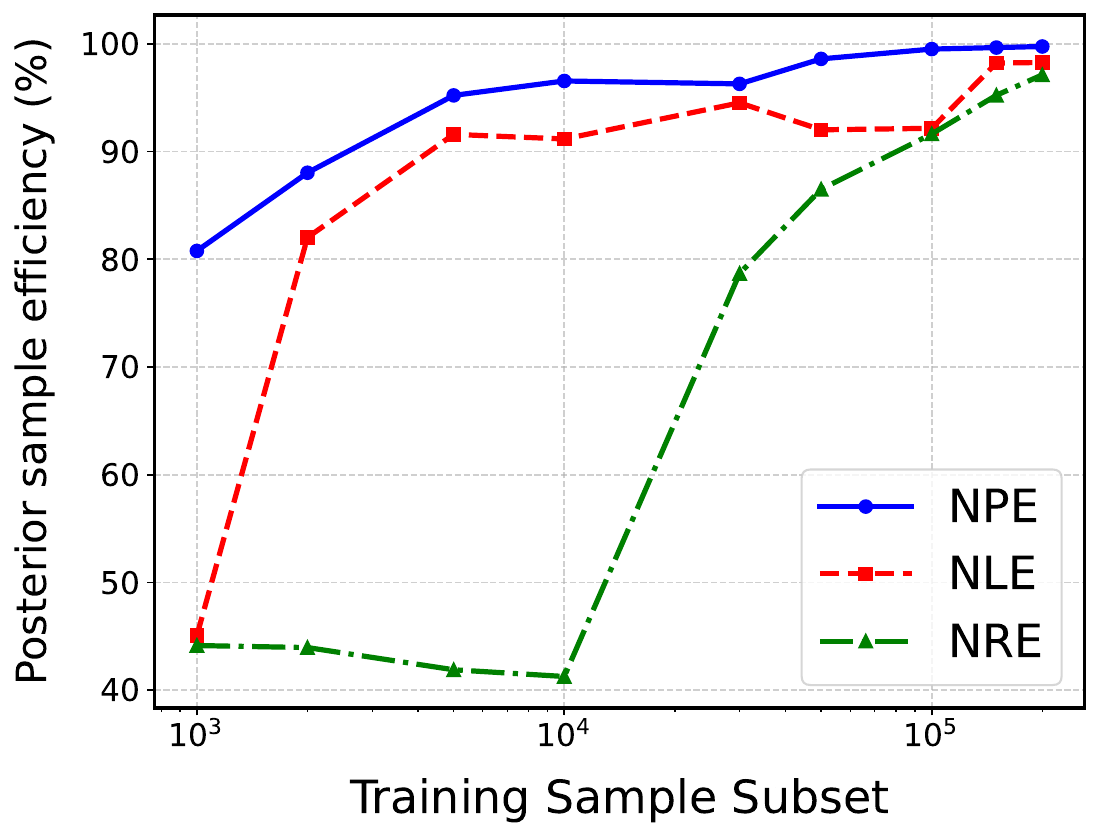} 
\includegraphics[width=0.49\textwidth]{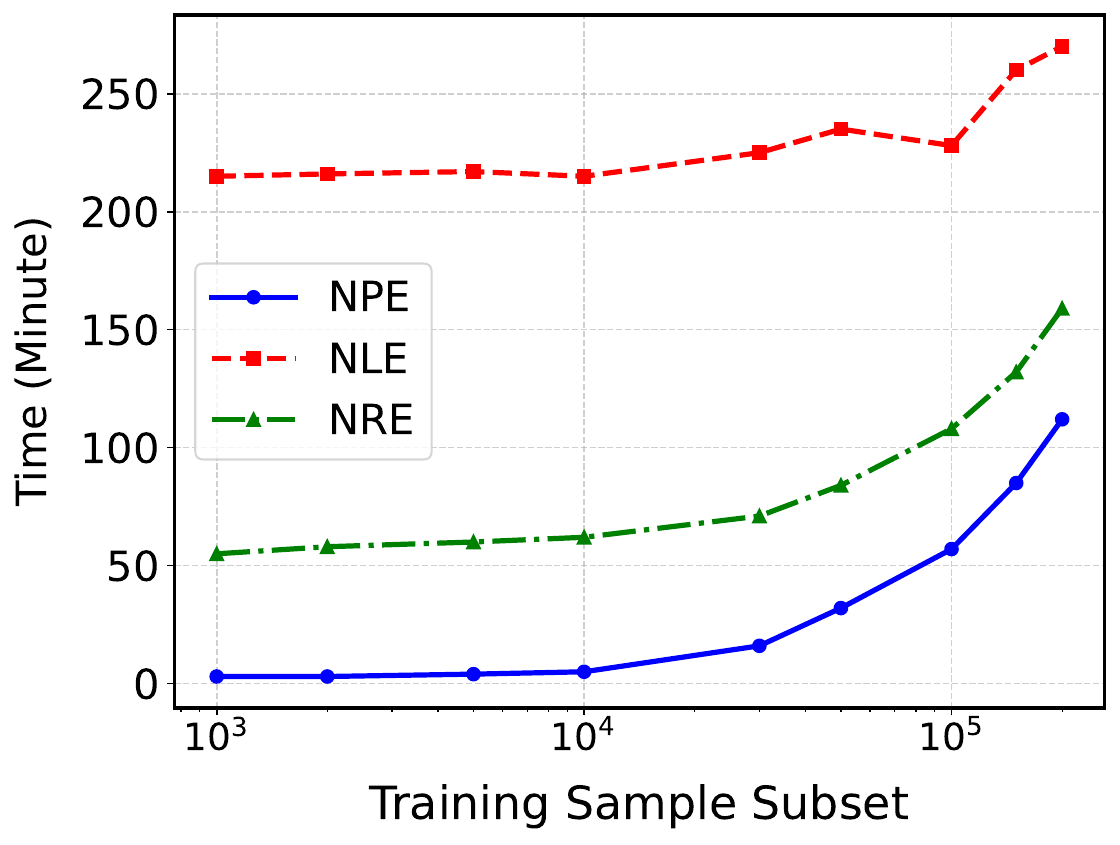}
\caption{The comparison of posterior sample efficiency (left) and the time taken for different sample subsets corresponding to three different algorithms is displayed here.}
\label{fig:com_eff}
\end{figure}

As mentioned in Section.~\ref{sec:data_file}, we are interested in finding the posterior efficiency of model parameters, defined as 
\begin{equation}
	\label{eq:eff}
	\text{Posterior efficiency} = \frac{\text{No of surviving samples within } 3\sigma \text{ range of observable}}{\text{Total number of posterior samples drawn}}
\end{equation}
   Figure~\ref{fig:com_eff} shows the posterior sample efficiency (left panel) and time consumption (right panel) for each SBI method. In both panels, the blue (solid), red (dashed), and green (dot-dashed) lines correspond to the NPE, NLE, and NRE methods, respectively.

From the left panel of the figure, we observe that posterior efficiency is not always directly proportional to the number of samples as also noted in Ref.~\cite{Morrison:2022vqe}. Notably, the NPE algorithm achieves higher posterior sample efficiency compared to NLE and NRE. Furthermore, the posterior sample efficiency for the NPE algorithm saturates after $1\times 10^5$ samples, whereas the NLE and NRE algorithms do not exhibit such behavior. To reach the desired posterior efficiency in our model, NPE requires only 50\% of the total sample set, whereas NLE and NRE require the full sample.

\begin{figure}[!htb]
\centering
\includegraphics[width=1.0\textwidth]{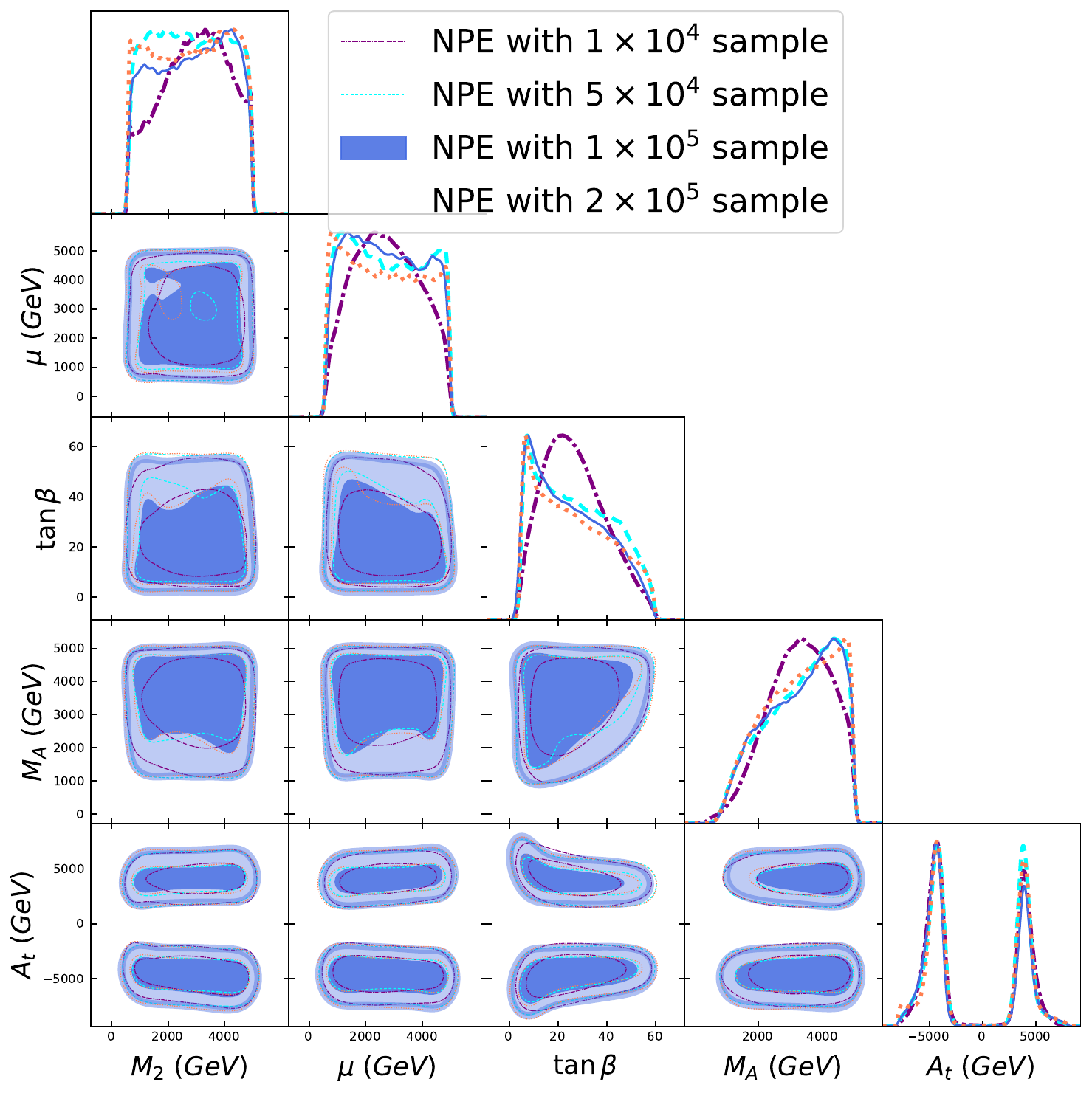}
\caption{2D and 1D Posterior distributions of all the parameters coming from NPE method with different sample subsets are shown here. The lighter and darker blue colors represent the $2\sigma$ and $1\sigma$ regions with $1\times 10^5$ sample subset. The sample subset contours are represented by purple, cyan, blue, and coral colors for $1\times 10^4$, $5\times 10^4$, $1\times 10^5$, and $2\times 10^5$ samples respectively.}
\label{fig:npe_posterior}
\end{figure}
Next, we focus on the time taken for training and posterior sample generation. The time consumption for each algorithm is presented in the right panel of Figure~\ref{fig:com_eff}. It is obvious that the NPE is significantly faster than the other two algorithms. 
 For instance, in the case of $1\times 10^5$ sample, the NLE (NRE) takes approximately 4 (2) times longer than the NPE. Even for the full sample set, while the posterior sample efficiency is nearly the same across all algorithms, the time consumption for the NLE and NRE remains substantially higher than that for the NPE.
  
Similar to the TARP test results, these observations on posterior sample efficiency and time consumption further confirm that the NPE algorithm provides superior performance compared to the other two algorithms, even with fewer samples. Based on this analysis, we conclude that the $1\times 10^5$ sample subset is sufficient to achieve the desired results using the NPE algorithm.

We finally derive the posterior distribution of the allowed parameter space for pMSSM model after satisfying the constraints from Higgs sector and flavor observables. The corresponding corner plots of posterior samples for the NPE algorithm are shown in Figure~\ref{fig:npe_posterior}. In this plot, purple, cyan, blue, and coral lines represent subsets with $1\times 10^4$,  $5\times 10^4$, $1\times 10^5$, and $2\times 10^5$ samples, respectively. Filled contour plot is shown only for the $1\times 10^5$ sample subset, while the other subsets are represented with solid or different kinds of dashed lines. 
It is evident from these plots that there is no significant distinction among the contour plots for subsets with $1\times 10^5$ samples or more. 

At the tree level, the two parameters that describe the MSSM higgs sector are $M_A$ and $\tan\beta$. At high $M_A$, which is also called the decoupling limit\cite{djouadi2008anatomy,Djouadi:2013lra}, the lighter CP-even state has properties close to that of the SM-like higgs. However, one requires significant loop correction to push the mass to the experimentally measured value. This loop correction is largely obtained from the third generation squarks, especially top squark. Therefore, the top squark mixing parameter $X_t = A_t - \mu\cot\beta$ plays a big role in determining the higgs mass. This implies a larger $\tan\beta\gtrsim 10$ and large $A_t$. This is exactly what we have obtained through our analysis. As evident from Figure~\ref{fig:npe_posterior}, the regions with $A_t$ close to zero are disfavored and we have obtained two distinct regions for $A_t$ beyond 1 TeV on either side of zero. Clearly, the NPE algorithm can efficiently understand this crucial relation between the $A_t$ parameter and the SM-like Higgs mass.      
The value of $\mu$ does not have much impact as long as $A_t$ dominates the mixing parameter $X_t$. This explains the posterior of $\mu$, which is mostly flat. The parameter $M_2$ can only impact the Higgs sector slightly through loop corrections, and hence, the corresponding distribution is also mostly flat.  

\subsection{$M_A-\tan\beta$ contour}
\label{sec:ma_tanbeta}
\begin{figure}[h!]
\centering
    \includegraphics[width=0.79\textwidth]{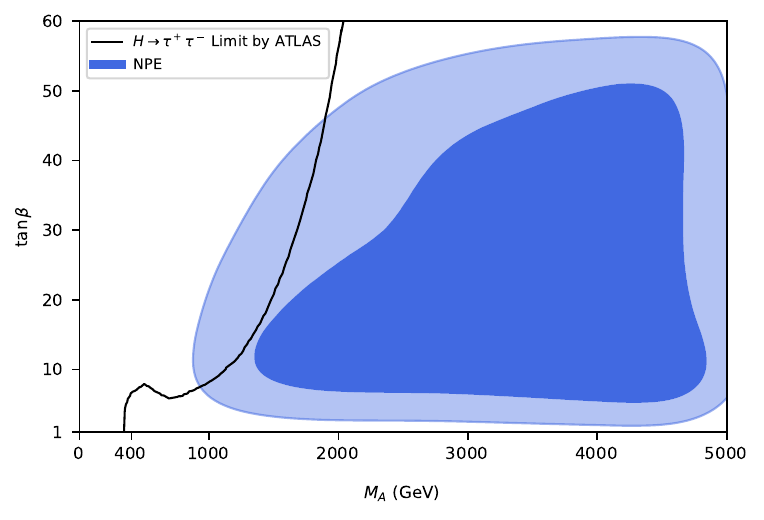}
    \caption{$M_A$-$\tan\beta$ contour plot for NPE method with $1\times 10^{5}$ sample subset is displayed where darker and lighter blue colors correspond to $1\sigma$ and $2\sigma$ regions respectively. The ATLAS limit ~\cite{ATLAS:2020zms} coming from $H\to \tau^+ \tau^-$ search is shown by a solid black line.}
    \label{fig:ma_tanbeta}
\end{figure}
In this section, we discuss the allowed region for $M_A-\tan\beta$ plane determined by the SBI methods. Figure~\ref{fig:ma_tanbeta} displays the $M_A-\tan\beta$ contour plot based on results from the $1\times 10^5$ sample subset. The lighter and darker blue colors represent the $2\sigma$ and $1\sigma$ contours respectively for NPE algorithm. 
The allowed region here crucially depends on the flavor observables as well as the higgs sector observables. Constraints arising from the ratio of $\mathcal{B}r(B\to\tau\nu)$ in MSSM and SM depend on the doubly charged higgs mass (determined by $M_A$) and $\tan\beta$. Similarly, $\mathcal{B}r(B_s \to \mu^+\mu^-)$ varies as $\sim\frac{\tan^6\beta}{M_A^4}$ for large $\tan\beta$. Clearly, a combination of large $\tan\beta$ and small $M_A$ should be disfavored from experimental observations. The $\mathcal{B}r(B\to X_s\gamma)$ also depends on $M_A$ since the charged Higgs appears at one loop contribution to the observable. When incorporating Higgs properties alongside these flavor observables, a similar pattern in the $M_A-\tan\beta$ plane emerges, consistent with the SBI methods, as shown in Figure~\ref{fig:ma_tanbeta}. The most stringent direct search constraint on the heavy CP-even higgs mass arises from the $H \rightarrow \tau^+\tau^-$ channel. The ATLAS Collaboration has provided the exclusion limit on Heavy Higgs based on the value of $\tan\beta$ in the Ref.~\cite{ATLAS:2020zms} using Run-II data. We have shown this exclusion limit by the black colored line on this contour plot in Figure.~\ref{fig:ma_tanbeta}. It is clear from the figure that the $1\sigma$ region obtained through the most efficient NPE method is allowed, but part of the 
$2\sigma$ allowed region is excluded by the experimental result for $M_A < $ 2000 GeV. This result is in good agreement with the results provided by Ref.~\cite{Barman:2024xlc}. 

\subsection{Performance Comparison: NPE Algorithm vs. MCMC Method}
\label{sec:com_npe_mcmc}
\begin{figure}[!htb]
\centering
\includegraphics[width=0.99\textwidth]{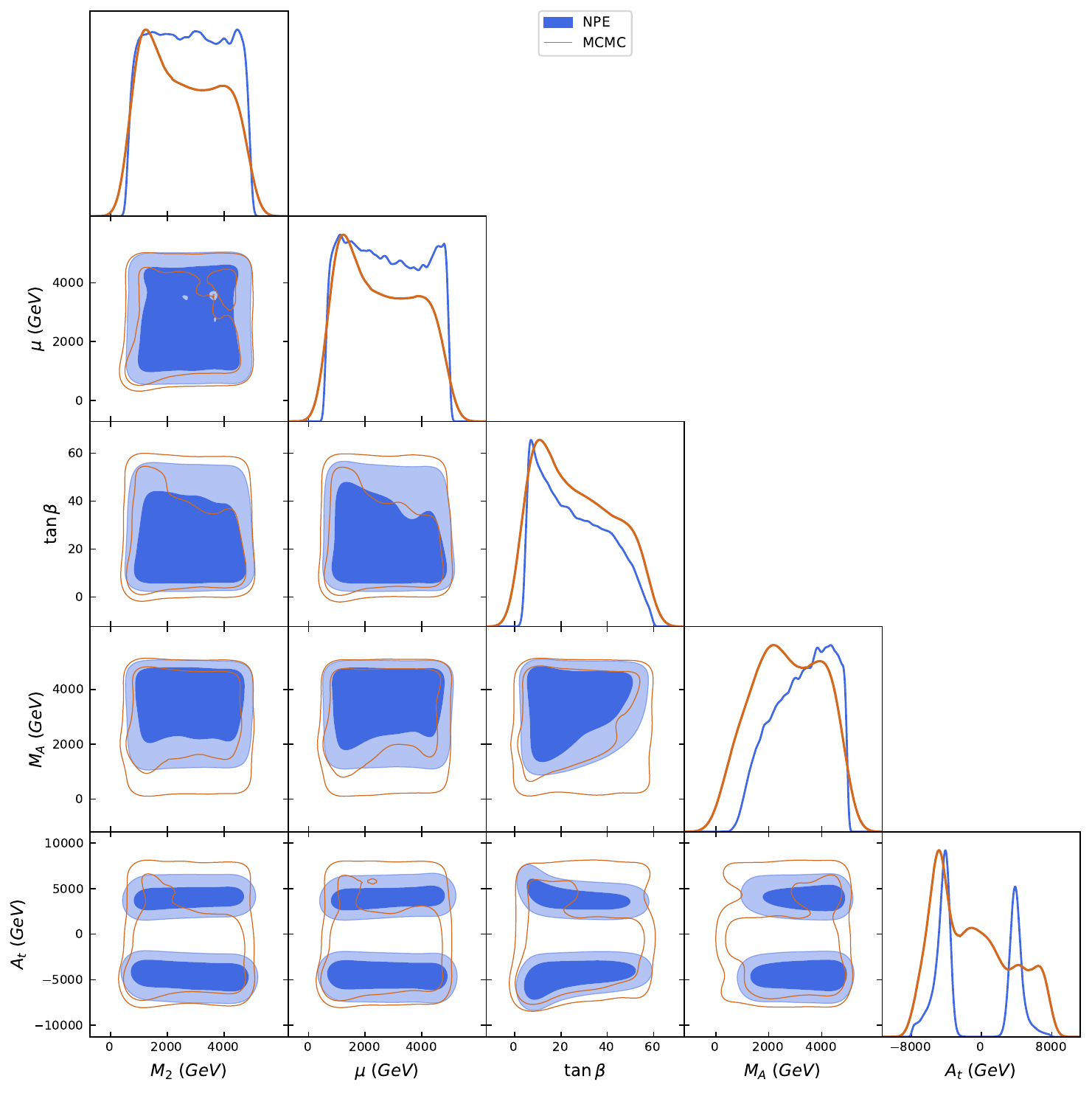} 
\caption{The comparison of the posterior distribution coming from NPE and MCMC methods is shown with blue contour and brown line respectively.}
\label{fig:com_npe_mcmc}
\end{figure}
We also perform the MCMC analysis using the same set of observables and parameter sets for this pMSSM5 model and compared the results to those obtained via the NPE method. To make the comparison as fair as possible, we use the same computational set-up as mentioned in the footnote of the Section~\ref{sec:data_file} (see footnote 2). We run the MCMC setup  with 4 cores, mirroring the parallelization in SBI approaches, using 500 walkers with 1000 points per walker. We discard the first 20\% of samples as burn-in to mitigate the influence of initial conditions. Interestingly, the MCMC run takes roughly 72 hours to complete, whereas the NPE method, using a subset of $1\times 10^5$ sample subset, completes in only 24 hours.

Also, we show the posterior distribution coming from each method in the Figure~\ref{fig:com_npe_mcmc}, where the blue colored contour refers to the NPE method and the brown colored line refers to the MCMC result. 
The posterior distribution of $A_t$ (see the first diagonal plot from right in the bottom row of Figure~\ref{fig:com_npe_mcmc})  
obtained from the SBI and MCMC analyses shows a significant disagreement around $A_t=0$.  The parameter $A_t$ plays an important role in the radiative corrections to $m_h$ and smaller values of $A_t$ are disfavored as they lead to $m_h<123$ GeV
(see e.g., Figure~1 of Ref.~\cite{Djouadi:2013lra} and Figure~7 of \cite{Christensen:2012ei}) 
While SBI method correctly captures this feature, the MCMC analysis fails to reproduce the dip around $A_t=0$. 
This discrepancy can be attributed to the chi-square evaluation within the MCMC framework: the derived value of one observable can differ from the central value provided by experiment, while others can fit closely to the experimental results, and we still get a lower chi-square value. 
Consequently, the MCMC framework tends to allow a slightly larger parameter space for many parameters like $M_A,\tan\beta$, etc. It is worth mentioning that low $M_A$ and low $tan\beta$ regions (third column in second row from bottom of Figure~\ref{fig:com_npe_mcmc}), allowed by the MCMC analysis, lie outside of SBI predicted 2$\sigma$ regions.  
As mentioned earlier, SBI predicted posterior distributions show good agreement with the results presented in Ref.~\cite{Barman:2024xlc} (see Figure~1).
We conclude that the NPE method performs better in terms of both computational efficiency and posterior accuracy. Although the computational time in the MCMC method can be reduced by considering a smaller number of walkers and fewer steps per walker, it should be noted that even with a large number of walkers and steps, the MCMC analysis fails to fully capture the true posterior distribution. Therefore, further reduction in these numbers is likely to yield an even less accurate posterior.


\section{LtU-ILI framework for pMSSM9 with Higgs, flavor and DM constraints}
\label{sec:dm}
In this section, we focus on exploring the region of pMSSM parameter space with nine free parameters (pMSSM9), 
consistent with Dark Matter (DM) constraints together with other observables such as the Higgs boson mass, Higgs coupling strengths to SM particles, and flavor-physics constraints, as discussed in Section~\ref{sec:observable}\footnote{A few phenomenological analyses in this context dark matter in pMSSM scenario can be seen in Refs.~\cite{Han:2013gba,Chakraborti:2017dpu, Chakraborti:2015mra, Chakraborti:2014gea, Choudhury:2013jpa, Choudhury:2012tc, Bhattacharyya:2011se}}. For the DM sector, we impose bounds on the DM relic density ($\Omega h^2$), the spin-independent (SI) nucleon-DM scattering cross section ($\sigma_{\text{SI}}$), and the spin-dependent (SD) cross sections for protons ($\sigma^p_{\text{SD}}$) and neutrons ($\sigma^n_{\text{SD}}$). Including the theoretical uncertainty~\cite{Baro:2007em, Baro:2009na, Banerjee:2021hal}, we have considered $\Omega h^2 = 0.12 \pm 0.01$ as considered in the Ref.~\cite{Morrison:2022vqe}\footnote{The DM relic density as provided by the {Planck} 2018 results is $\Omega h^2 = 0.12 \pm 0.001$~\cite{Planck:2018vyg}.}. 
The upper limits on the spin-independent (SI) nucleon-DM scattering cross section ($\sigma_{\text{SI}}$) depend on the DM mass. 
For example, for 100 GeV WIMP are approximately $9.1\times10^{-47}$ cm$^2$ and $4.4\times10^{-47}$ cm$^2$, as reported by \textbf{XENON1T}~\cite{XENON:2018voc} and \textbf{XENONnT}~\cite{XENON:2023cxc}, respectively.
For the spin-dependent proton cross-section, the upper limits on $\sigma^p_{\text{SD}}$ are $4.2\times10^{-41}$ cm$^2$ and $3.29\times10^{-40}$ cm$^2$ corresponding to 100 GeV and 1000 GeV DM masses respectively coming from \textbf{PICO-60}~\cite{PICO:2019vsc} experimental result. Similarly, the upper limit on spin-dependent neutron cross-section ($\sigma^n_{\text{SD}}$) according to \textbf{PandaX-4T}~\cite{PandaX:2022xas} experimental result are $8.7\times10^{-42}$ cm$^2$ and $6.26\times10^{-41}$ cm$^2$ for 100 GeV and 1000 GeV DM masses respectively. We implement the bin-wise upper bounds on the cross-sections provided by the \textbf{XENON1T}, \textbf{PandaX-4T}, and \textbf{PICO-60} experiments.

\begin{table}[!htb]
\begin{center}
\begin{tabular}{||c|c||}
\hline\hline
\textbf{Parameter} & \textbf{Range} \\ 
\hline\hline
$M_1$ & 100-2000\\
\hline
 $M_2$ & 100-5000 \\
 \hline
 $M_3$ & 2500-5000 \\
 \hline
 $\mu$ & 100-5000 \\
 \hline
 $\tan\beta$ & 1-60 \\
 \hline
 $M_A$ & 100-5000 \\
 \hline
 $|A_t|$ & 0-8000 \\
 \hline
 $m_{\tilde{l}}$ & 1000-5000 \\
 \hline
 $m_{\tilde{q}}$ & 2000-5000 \\
\hline\hline
\end{tabular}
\caption{This table represents the range of all nine free parameters considered for our analysis. All the parameters mentioned here have unit GeV except $\tan\beta$. 
}
\label{tab:parameter_dm}
\end{center}
\end{table} 

The parameter space considered corresponding to pMSSM9 for the sampling is displayed in the Table~\ref{tab:parameter_dm}. Here, the left- and right-handed sleptons for all three generations have the same mass. Similarly, all three generations left- and right-handed squarks also have the same mass. The parameters other than these 9 mentioned in Table~\ref{tab:parameter_dm} are fixed at the same value as mentioned in Section~\ref{sec:parameter}. Also, the Higgs mass, coupling strength, and flavor physics observables are considered with the same values as mentioned in Section~\ref{sec:observable}.

\begin{figure}[!htb]
\centering
\includegraphics[width=0.3\textwidth]{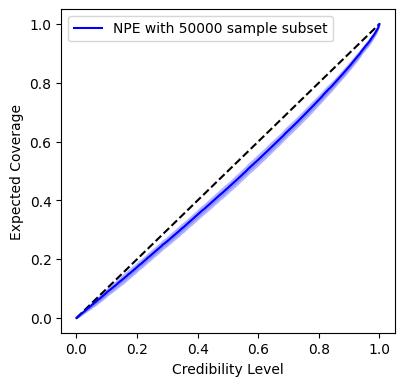}
\includegraphics[width=0.3\textwidth]{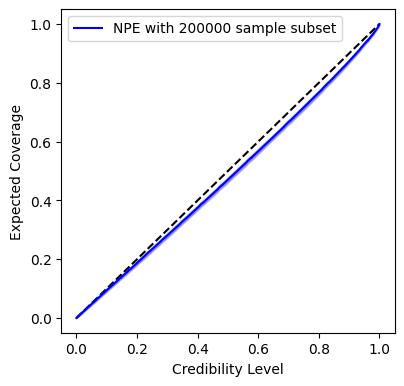}
\includegraphics[width=0.3\textwidth]{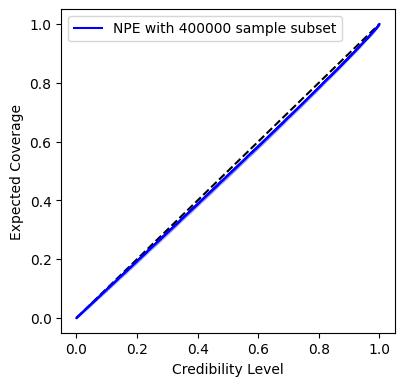}\\
\includegraphics[width=0.3\textwidth]{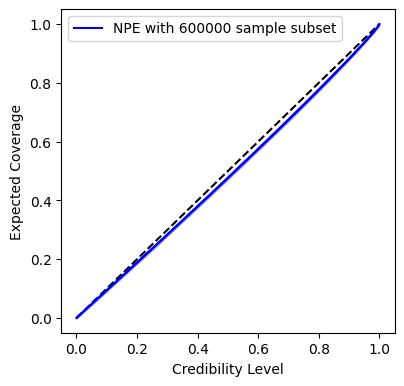}
\includegraphics[width=0.3\textwidth]{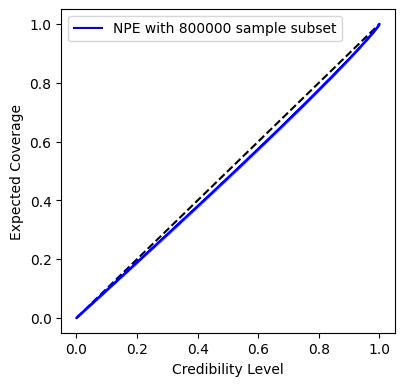}
\includegraphics[width=0.3\textwidth]{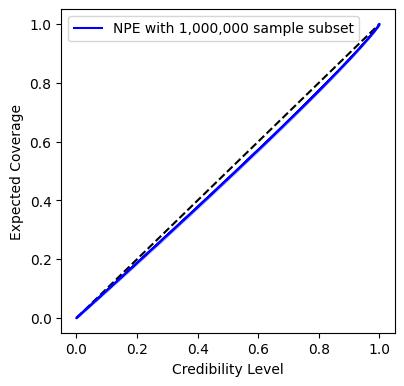}
\caption{TARP test plot corresponding to $5\times 10^4$, $2\times10^5$, $4\times10^5$, $6\times 10^5$, $8\times10^5$, $1\times10^6$ sample subsets with NPE method are displayed here. }
    \label{fig:tarp_dm}
\end{figure}

Here, we focus on the fact that $\lspone$ is the Weakly Interacting Massive Particle (WIMP) DM, and this condition has been imposed during the sample preparation before applying the SBI method. We have generated around $1\times 10^{6}$ samples allowing for Higgs and flavor data within their respective $3\sigma$ ranges as discussed in Section~\ref{sec:data_file}. At this stage, we did not apply any DM  constraints\footnote{Inclusion of DM constraints at this stage makes it very difficult to find valid parameter points and as a result the computation becomes extremely time consuming~\cite{Morrison:2022vqe}}. In this $1\times 10^{6}$ point sample, there are about $\sim 2\times 10^3$ points which satisfy all the conditions including DM. We have observed, as discussed subsequently, that using the SBI method, mainly NPE, we can effectively increase this efficiency. First, we have shown the TARP plots for different sample subsets in Figure~\ref{fig:tarp_dm}. It shows that after $4\times 10^5$ sample subset, the blue band closely matches with the diagonal line.
\begin{figure}[!htb]
\centering
\includegraphics[width=0.7\textwidth]{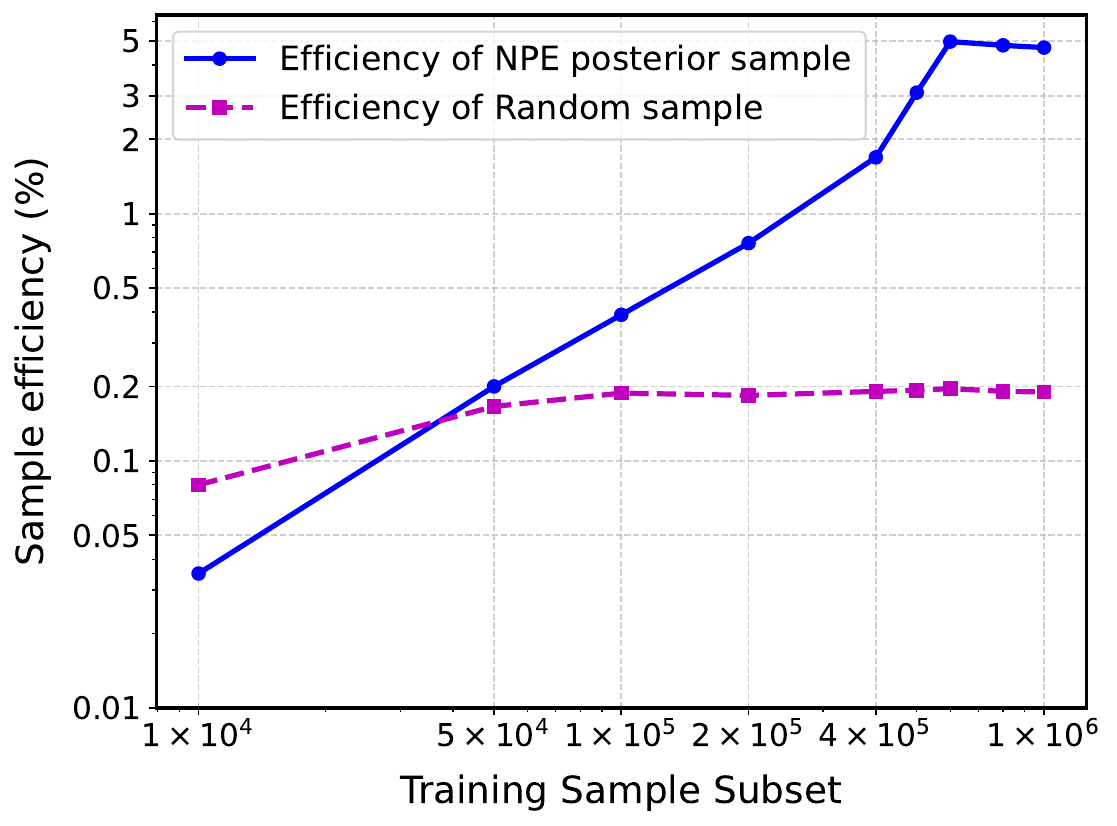} 
\caption{The efficiencies of NPE posterior sample and random sample are shown for different sample subset where the blue solid and magenta dashed colored lines refer to the posterior sample and random sample, respectively.}
\label{fig:com_eff_dm}
\end{figure}
In Figure~\ref{fig:com_eff_dm}, we compare the efficiencies of the random sample and the posterior sample for different subsets. Here, the blue solid line corresponds to the efficiency of the NPE posterior sample and the magenta dashed line corresponds to that obtained from random sample. From this figure, it is clear that NPE provides much better efficiency than the random sample when the training sample is sizable. The NPE method provides the maximum of $\sim$5.0\% sample efficiency whereas the maximum efficiency for the random sample with the same sample size is $\sim$0.2\%. NPE is 25 times more efficient than the random sampling method which also highlights the advantage of using the NPE method for such models with a wide variety of observables. 
Note that the efficiency obtained from NPE method is much less as compared to the results shown earlier for the scenario without including DM observables discussed in Section~\ref{sec:posterior_eff}. This is expected since the training sample had only a handful of points that satisfied the DM constraints. Naturally, the algorithm here is not well-trained on the DM allowed parameter space. With better computational resources, one can populate the training sample with more DM allowed points and enhance this efficiency factor.


\subsection{Impact of DM constraints on available parameter space}
\label{sec:dm_impact}
We now consider the posterior samples that satisfy all the constraints from Higgs and flavor observables.  
The impact of DM constraints is examined by imposing the observed relic density and the limits from direct-indirect detection experiments. Finally, we present the results obtained after applying all the constraints simultaneously.
\begin{figure}[!htb]
\centering
\includegraphics[width=0.32\textwidth]{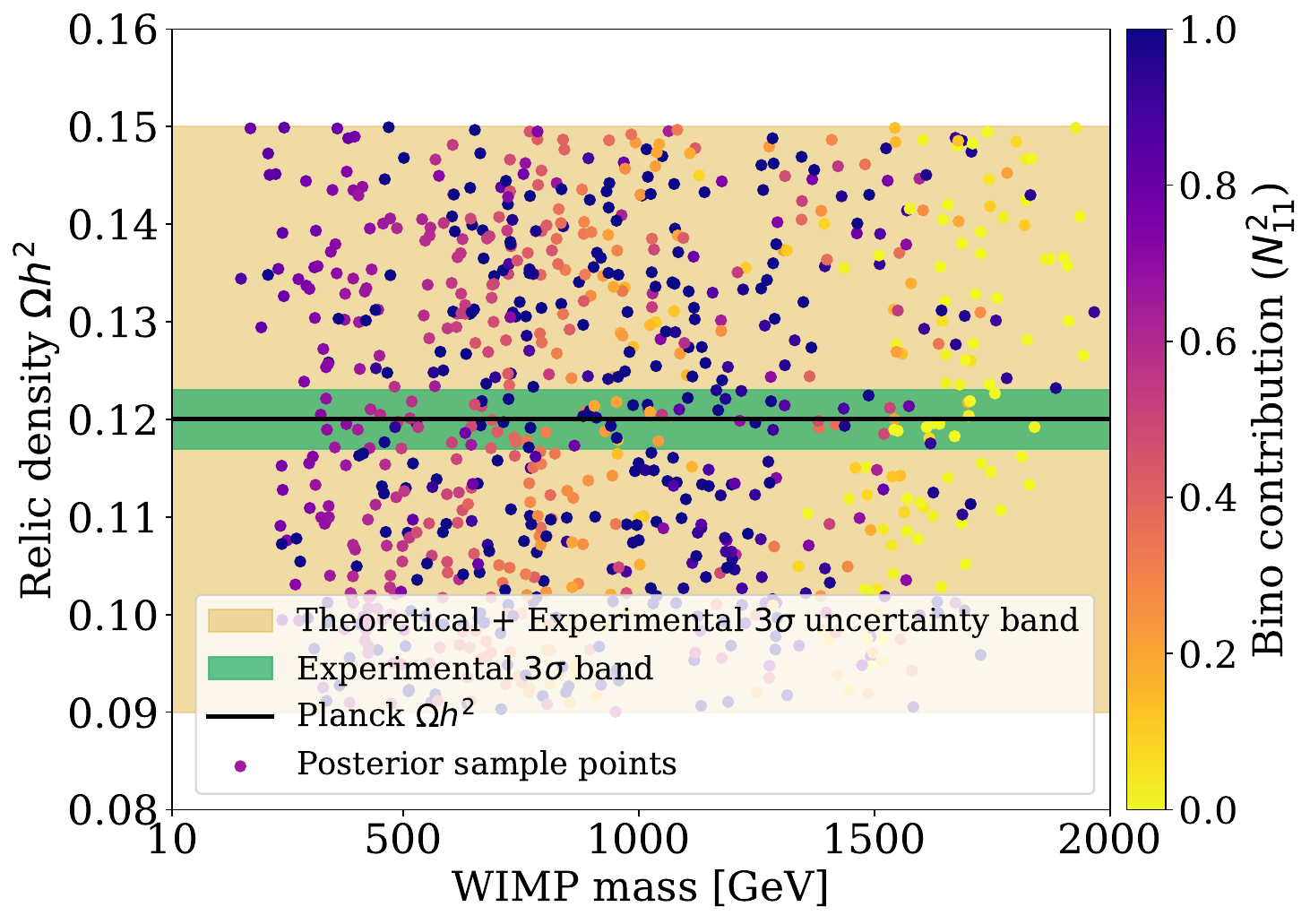} 
\includegraphics[width=0.32\textwidth]{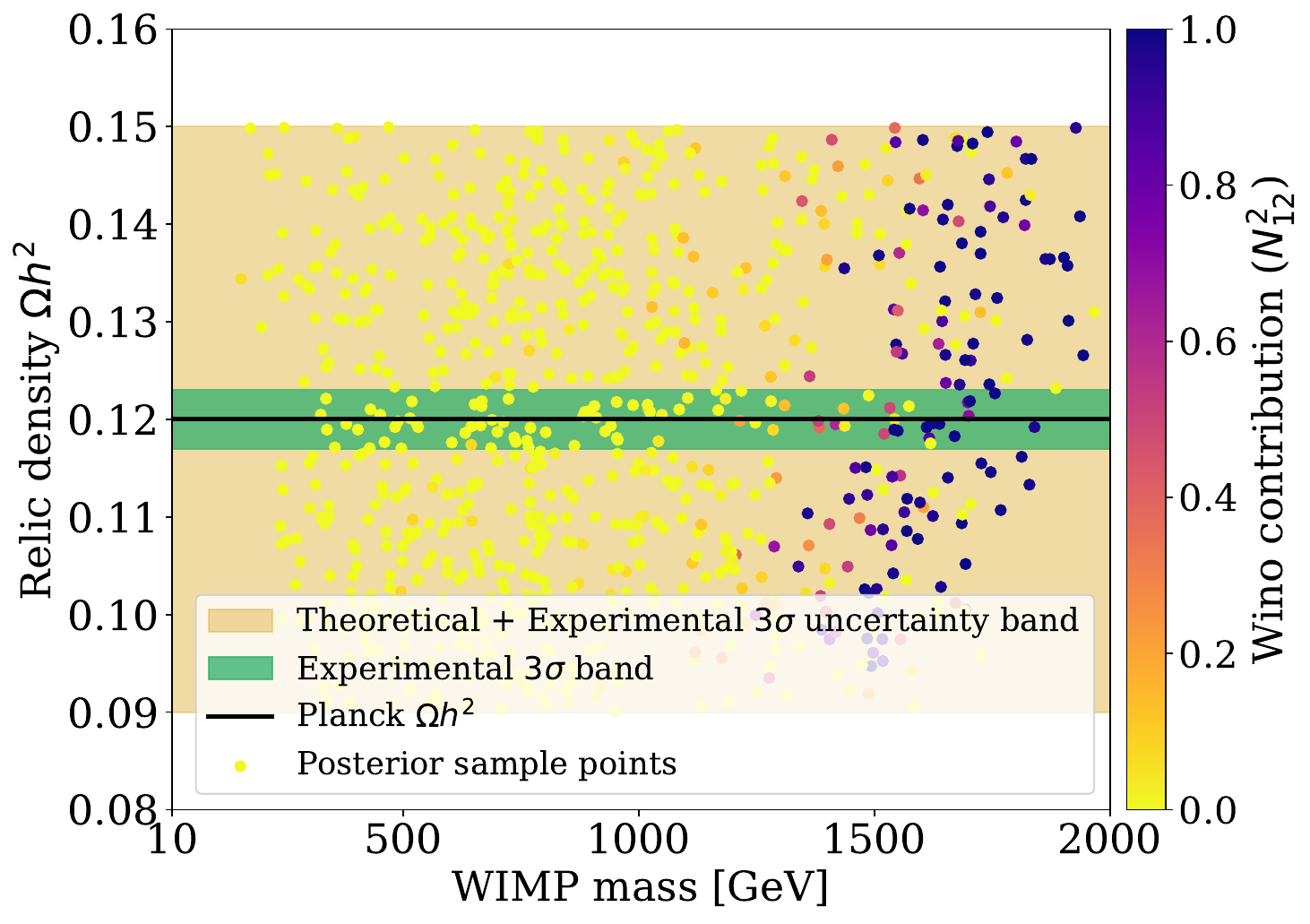}
\includegraphics[width=0.32\textwidth]{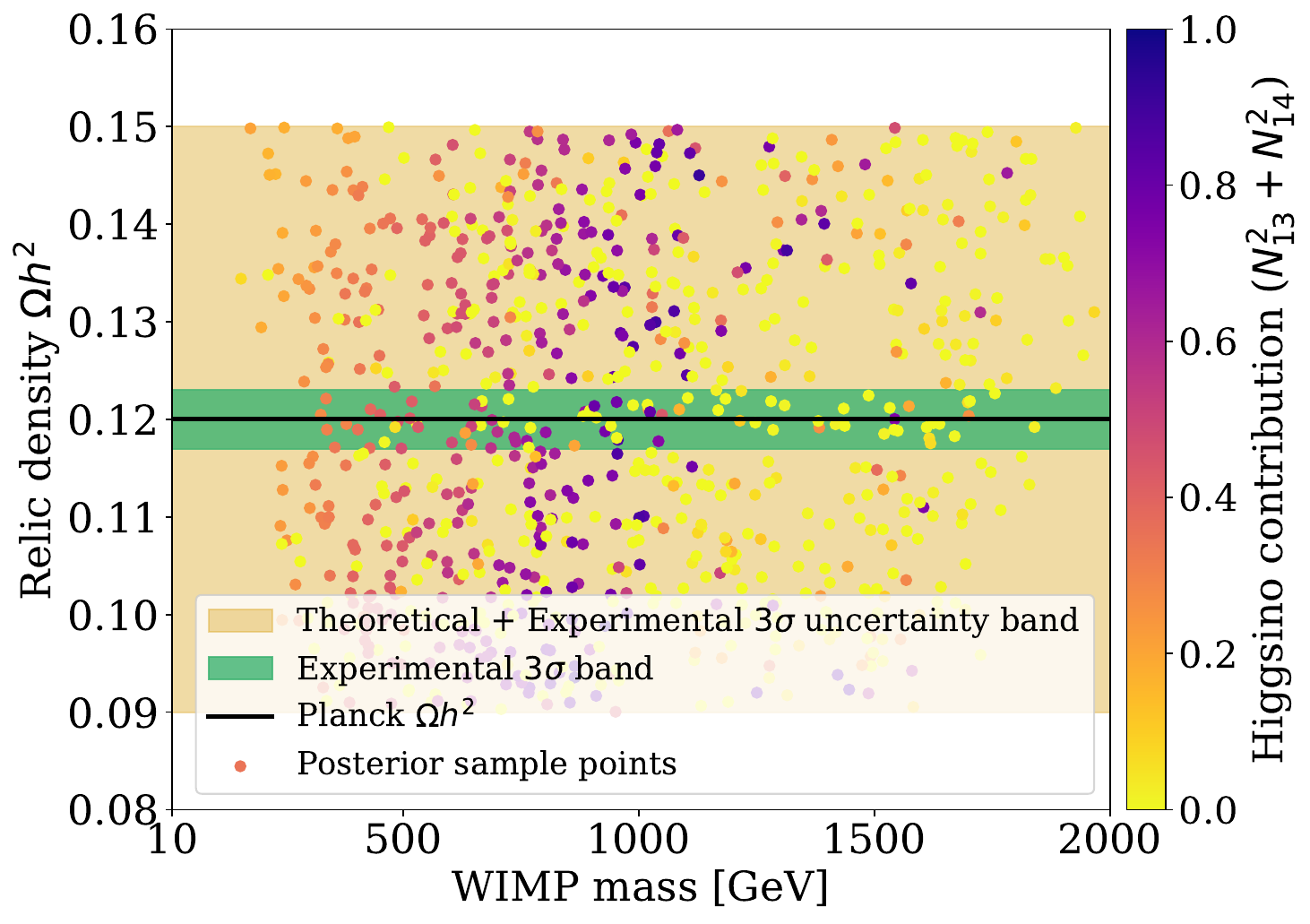}
\caption{The central value of observed relic density \cite{Planck:2018vyg}, experimental $3\sigma$ uncertainty band, and $3\sigma$ band of our consideration are shown with black solid line, green shaded region, and orange shaded region respectively. The color bar represents the bino, wino and higgsino admixtures 
in the LSP in left, middle, and right figures, respectively. }
\label{fig:omega}
\end{figure}
In Fig.~\ref{fig:omega}, we present the DM relic density as a function of the DM mass for all the points in the posterior sample obtained through the NPE method.
 The color coding in the three subfigures represents bino, wino and higgsino admixtures in the LSP respectively. Evidently, there is no real pattern arising from the bino-dominated LSP scenario.  Most points lying within the experimental $3\sigma$ relic density band are results of co-annihilation of the DM except those for which the LSP mass falls in the heavy Higgs (H/A) funnel regions, i.e., the LSP mass is nearly equal to half of the heavy Higgs mass. We observed that the co-annihilation occurs with one of chargino, stau or stop quark. The SBI predicted acceptable wino-dominated DM scenario lies only in the high mass range $\gtrsim 1500$ GeV. The higgsino-dominated DM scenario, on the other hand is only relevant close to the TeV range. These results are consistent with existing literature \cite{Han:2013gba,Chakraborti:2017dpu}. 

\begin{figure}[!htb]
\centering
\includegraphics[width=0.32\textwidth]{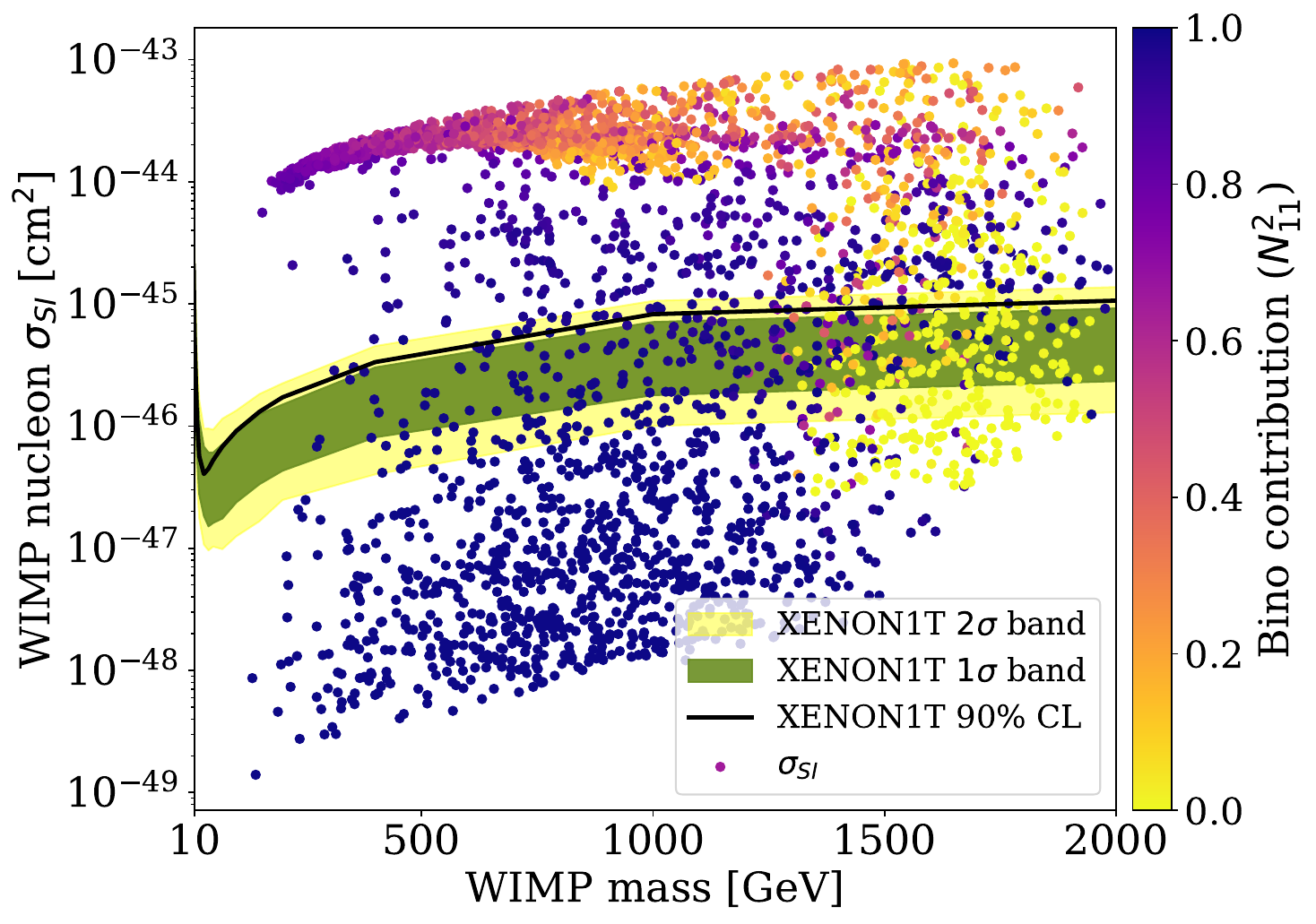} 
\includegraphics[width=0.32\textwidth]{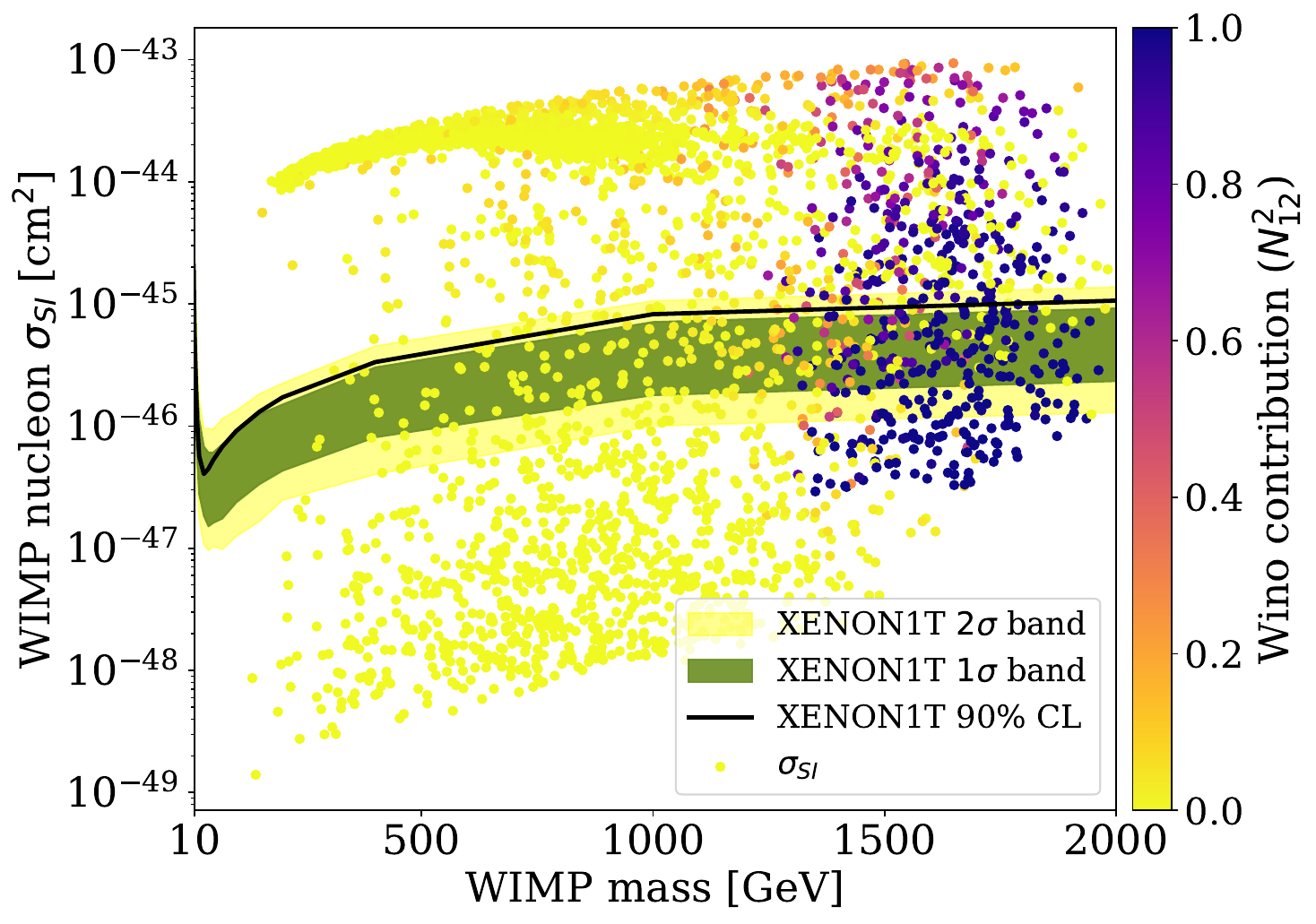}
\includegraphics[width=0.32\textwidth]{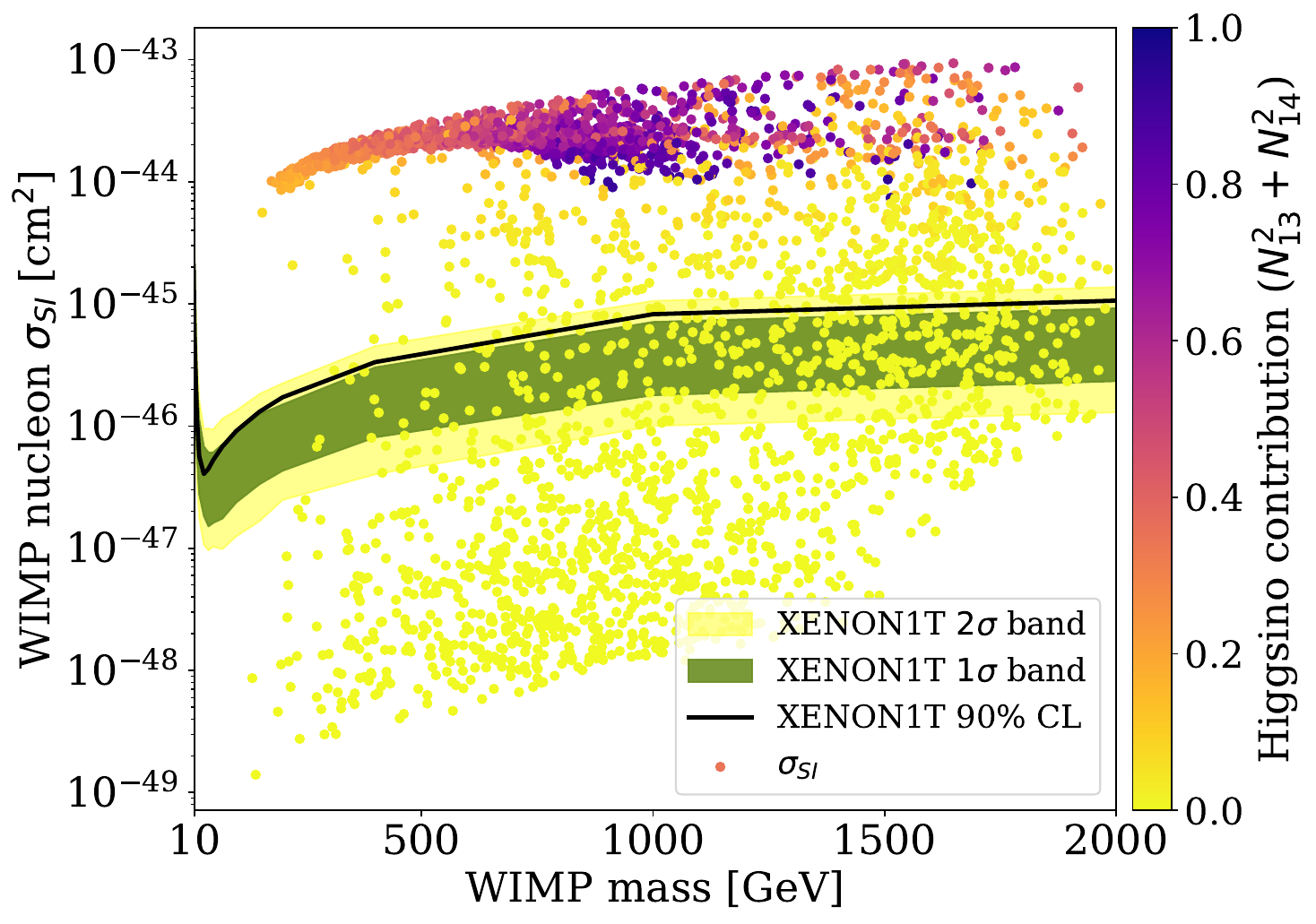}
\caption{The upper limit on the WIMP spin-independent nucleon cross-section coming from \textbf{XENON1T} experiment \cite{XENON:2018voc} and the $1\sigma$ and $2\sigma$ uncertainties are shown by black solid line, green shaded region, and yellow shaded region respectively. The scatter points refer to the cross-section   
 obtained from the posterior sample which satisfies the DM relic density condition along with Higgs and flavor constraints. The color bar represents the bino, wino, and higgsino contributions to the LSP in the left, middle, and right figures, respectively.}
\label{fig:xenon1t_sigma}
\end{figure}
\begin{figure}[!htb]
\centering
\includegraphics[width=0.32\textwidth]{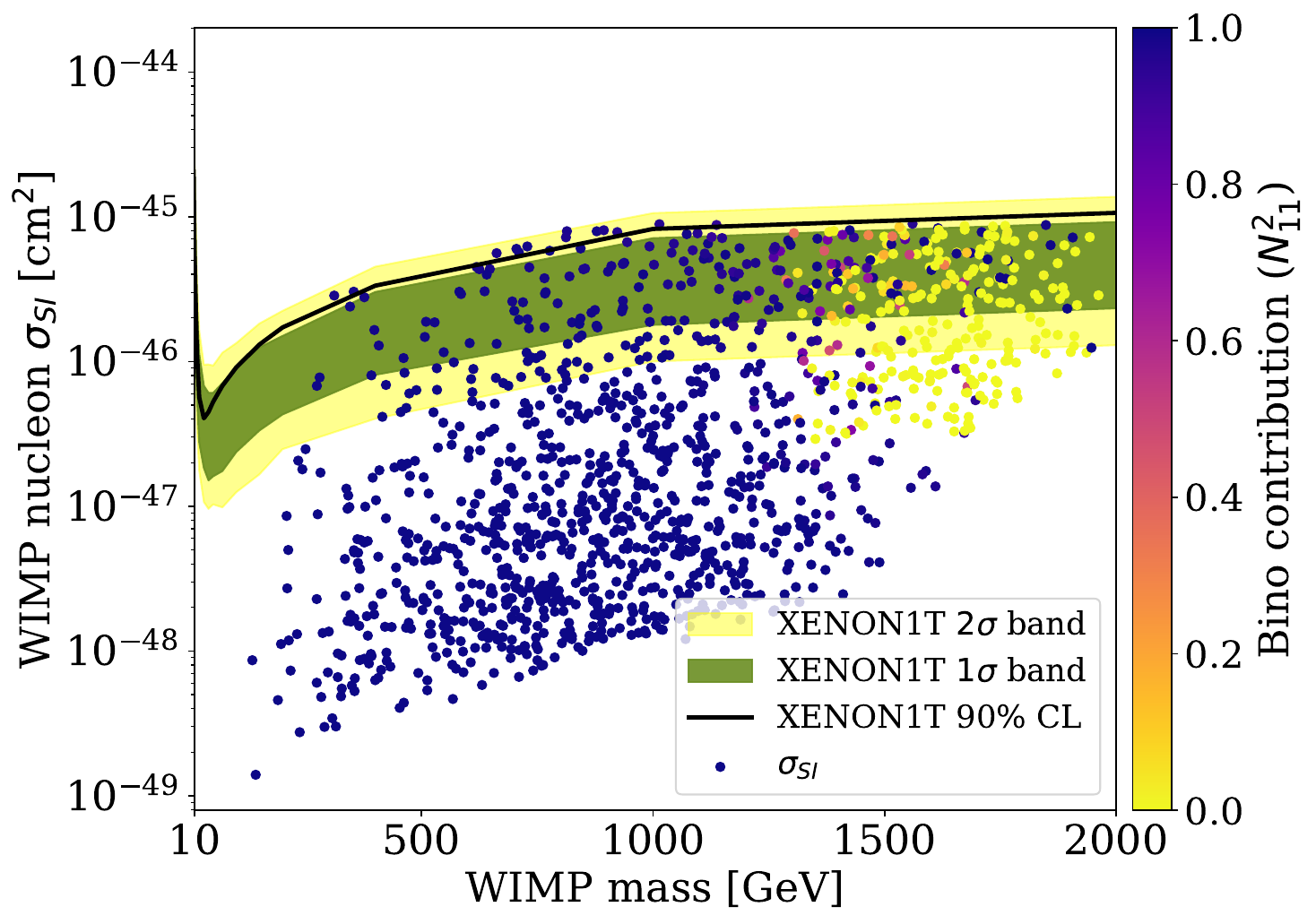} 
\includegraphics[width=0.32\textwidth]{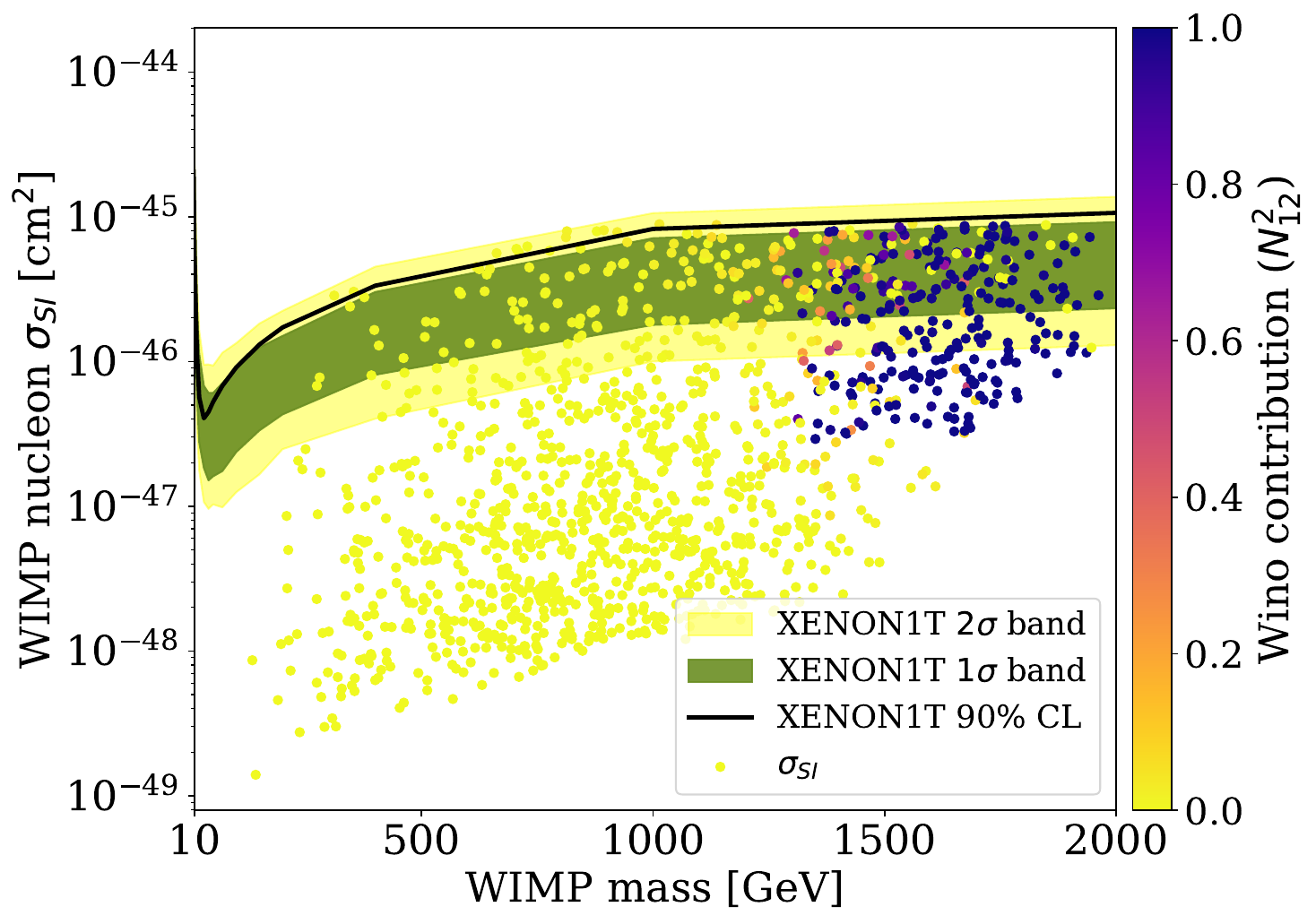}
\includegraphics[width=0.32\textwidth]{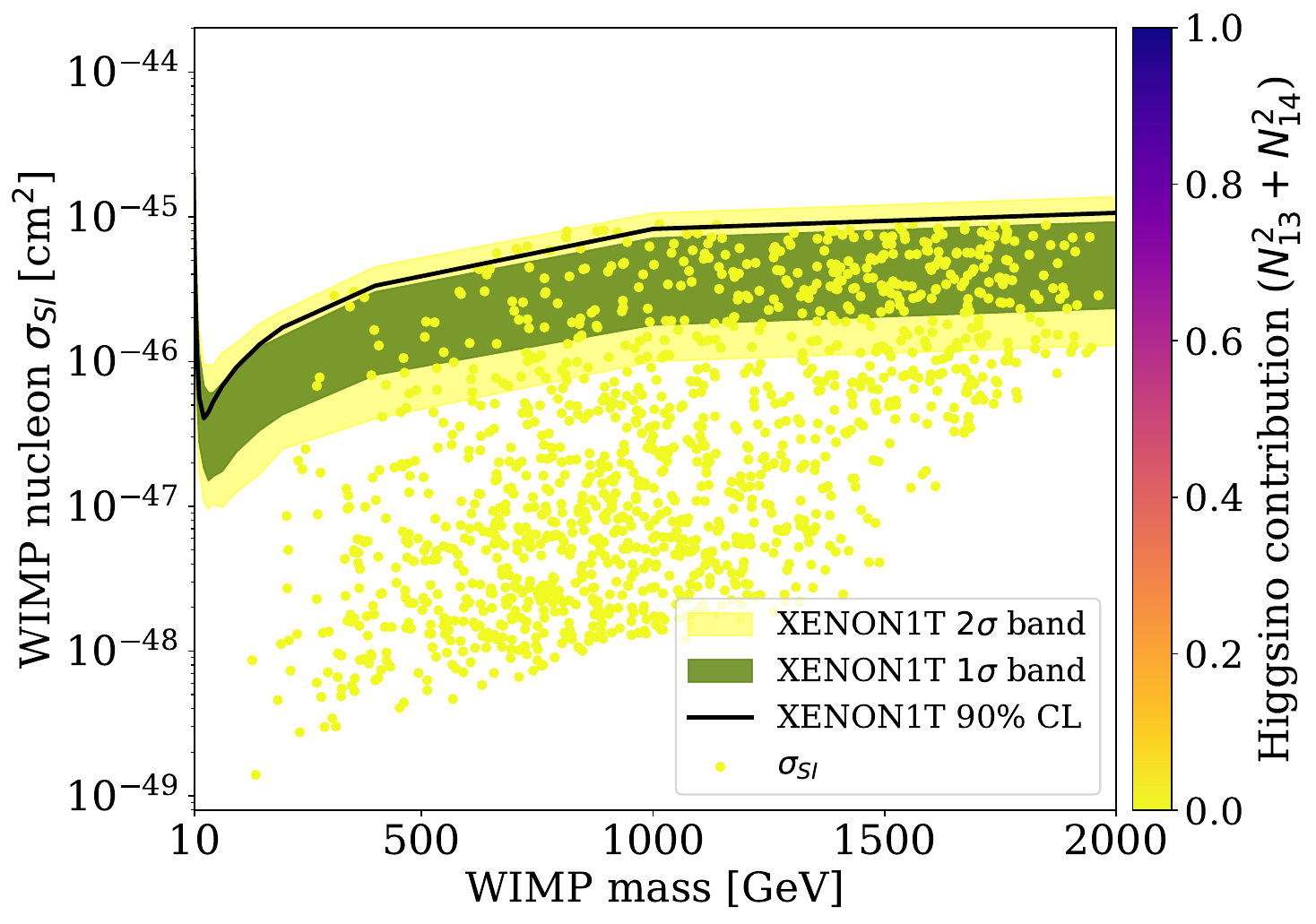}
\caption{The upper limit on the WIMP spin-independent nucleon cross-section coming from the \textbf{XENON1T} experiment and the $1\sigma$ and $2\sigma$ uncertainties are shown by black solid line, green shaded region, and yellow shaded region respectively. The scatter points refer to the cross-section obtained from the posterior sample which satisfies all the constraints. The color bar represents the bino, wino, and higgsino contributions to the LSP in the left, middle, and right figures, respectively.}
\label{fig:xenon1t_sigma_all}
\end{figure}

Figure~\ref{fig:xenon1t_sigma} showcases the direct detection cross-section calculation for all points as a function of the DM mass. Evidently, 
the higgsino-dominated scenario is completely ruled out by the \textbf{XENON1T} data while the wino-dominated scenario is allowed only in the higher mass range, i.e., for DM masses in excess of 1500 GeV. The bino-dominated DM scenario is still allowed in the low mass region. This pattern can be more clearly understood from Figure~\ref{fig:xenon1t_sigma_all}. In this figure, we  display only those points satisfying all constraints. There is a stark difference in the DM type in the low and high mass range. While WIMP masses up to $\sim 1500$ GeV are allowed from the \textbf{XENON1T} data only if they are bino-dominated, beyond this point the good points are dominantly wino. There are no viable points where the DM candidate is higgsino-dominated. This is because with heavy squarks, the Higgs exchange diagrams contribute the most in direct detection cross-section and large gaugino-higgsino mixing results in a large cross-section \cite{Chakraborti:2017dpu,Hisano:2004pv,Chattopadhyay:2010vp,Drees:1993bu}. In Appendix~\ref{sec:appendix} we present five benchmark points representative of SBI-predicted DM favored parameter space. 

\section{Summary}
\label{sec:summary}

In this work, we have addressed a very pertinent issue of parameter space sampling. With large number of input parameters and precisely measured observables it often gets very difficult to identify the most favored parameter space given a particle physics model. Tools like MCMC and nested sampling are efficient but extremely time consuming. The main bottleneck of these algorithms is likelihood computation at each point. In this work, therefore, we have explored the possibility of drawing likelihood-free inference using ML techniques. SBI is one such method that has gained attention in recent times. A subclass of SBI, called amortized, is of particular interest owing to the fact that once properly trained, these methods can generate posterior distributions without getting trained on new data.  We have explored three different methods, namely, NPE, NLE, and NRE. In order to ascertain whether these methods can faithfully generate the true posterior distributions of the new physics parameters, we perform the TARP test which can efficiently detect inaccurate inferences. To test this framework, we take a very well-known BSM scenario, namely, the pMSSM. As for observables, we consider the SM-like Higgs mass and signal strengths alongside the flavor observables crucial for constraining the pMSSM parameter space. We observe that the NPE method can faithfully generate the posterior distributions of the input parameters, whereas the NLE and NRE methods fail the TARP test. We have also compared results obtained with training data sets of different sizes and show that with as low as $5\times 10^4$ data points, one can generate faithful posterior distributions using the NPE method. It is worth mentioning that we chose to vary only a small set of input parameters for this analysis while keeping all other input parameters fixed at reasonable values that lie beyond the present experimental sensitivity. 

We proceed further to compare the performance of the amortized SBI method with that of MCMC. Our results clearly show that NPE method can generate more faithful posterior distribution with reasonable efficiency in roughly one-third of the time the traditional MCMC method takes to converge. While this analysis was performed with a simpler 5 parameter pMSSM, we further test the NPE method on a more complicated 9 parameter pMSSM scenario after including dark matter observables to our study in addition to the Higgs and flavor observables. Adding DM observables constrains the parameter space severely. Even then, the NPE method is able to generate reliable posterior distributions. In SBI predicted parameter space, the points satisfying DM constraints in addition to all other constraints are bino, wino-dominated or bino-wino admixture. Note that we explored LSP masses upto 2 TeV. We observed that the `good' bino dominated points mostly populate the region upto $\sim$ 1.5 TeV and they satisfy relic density constraint either through chargino/stop/stau co-annihilation or via heavy Higgs (H/A) funnel. Beyond this point and upto 2 TeV most of the predicted `good' points are wino-dominated. For these points, the relic density constraint is  satisfied  mainly through substantial chargino co-annihilation and in some cases with small contributions from stau co-annihilation as well.

\section*{Acknowledgement}
\label{sec:ack}
\noindent
A. Choudhury and S. Mondal acknowledge ANRF India for Core Research Grant no. CRG/2023/008570. A. Mondal acknowledges ANRF India for the financial support through Core Research Grant No. CRG/2023/008570. S. Mondal acknowledges ANRF India for Core Research Grant no. CRG/2022/003208.     

\newpage
\appendix
\section{Benchmark points from SBI predicted pMSSM9 parameter space}
\label{sec:appendix}
\begin{table}[!htb]
\begin{center}
\begin{tabular}{||c|c|c|c|c|c||}
\hline\hline
Parameter & \textbf{BP1} & \textbf{BP2} & \textbf{BP3} & \textbf{BP4} & \textbf{BP5} \\
\hline\hline
$M_1$ (GeV) & 1104 & 1220 & 1947 & 1337 & 1708\\
\hline
$M_2$ (GeV) & 1124 & 4561 & 4584 & 3803 & 1682 \\
\hline
$M_3$ (GeV) & 3084 & 4739 & 2880 & 3166 & 4474 \\
\hline
$\mu$ (GeV) & 2188 & 4364 & 2870 & 1796 & 4585\\
\hline
$M_A$ (GeV) & 2714 & 4072 & 2552 & 2693 & 3378\\
\hline
$\tan\beta$ & 33.96 & 52.08 & 29.28 & 46.17 & 35.90 \\
\hline
$M_{\tilde{l}}$ (GeV) & 3544 & 1387 & 2540 & 1631 & 1794 \\
\hline
$M_{\tilde{q}}$ (GeV) & 4029 & 3034 & 2122 & 4745 & 3976 \\
\hline
$A_t$ (GeV) & -3332 & -4599 & 4041 & -4948 & -5013 \\
\hline
$m_{\lspone}$ (GeV) & 1104 & 1220 & 1946 & 1335 & 1681 \\
\hline
$m_{\lsptwo}$ (GeV) & 1122 & 4349 & 2867 & 1796 & 1708\\
\hline
$m_{\chonepm}$ (GeV) & 1122 & 4349 & 2867 & 1795 & 1681\\
\hline
$m_{\tilde{\tau}_1}$ (GeV) & 3545 & 1233 & 2511 & 1585 & 1711\\
\hline
$m_{\tilde{t}_1}$ (GeV) & 4028 & 3033 & 1961 & 4745 & 3975 \\
\hline
$m_H = m_A$ (GeV) & 2714 & 4071 & 2552 & 2692 & 3378 \\
\hline
$N_{11}^2$ & 0.994 & 1.0 & 0.999 & 0.995 & 0.0005\\
\hline
$N_{12}^2$  &0.005 & 0.0 & 0.0 & 0.0 & 0.999\\
\hline
$N_{13}^2+N_{14}^2$  & 0.001 & 0.0 & 0.001 & 0.005 & 0.0005\\
\hline
$m_h$ (GeV) & 124 & 124 & 125 & 125 & 126 \\
\hline
$\mathcal{B}r(B \rightarrow X_s \gamma)~\times 10^{-4}$ & 3.27 & 3.20 & 3.38 & 3.25 & 3.25 \\
\hline 
$\mathcal{B}r(B_s \rightarrow \mu^+ \mu^-)~\times 10^{-9}$ & 3.16 & 3.75 & 2.65 & 3.36 & 3.29 \\
\hline 
$\frac{\mathcal{B}r(B \rightarrow \tau \nu)}{\mathcal{B}r(B \rightarrow \tau \nu)_{SM}}$ & 0.991 & 0.993 & 0.994 & 0.983 & 0.994 \\
\hline 
$\Omega h^2$ & 0.122 & 0.133 & 0.132 & 0.143 & 0.138 \\
\hline
$\sigma_{SI}$ ($\times 10^{-46}$cm$^2$) & 0.61 & 0.02 & 1.24 & 2.99 & 0.42\\
\hline
$\sigma_{SD}^{p}$ ($\times 10^{-44}$cm$^2$) & 1.21 & 0.07 & 19.1 & 6.03 & 0.24\\
\hline
$\sigma_{SD}^{n}$ ($\times 10^{-45}$cm$^2$) & 12.2 & 0.02 & 6.45 & 50.6 & 8.03 \\
\hline\hline
\end{tabular}
\caption{Benchmark points (\textbf{BPs}) representative of the SBI-predicted allowed pMSSM9 parameter space satisfying all constraints.
}
\label{tab:bp_dm}
\end{center}
\end{table}

\newpage
\bibliography{reference}

\end{document}